\documentstyle [eqsecnum,preprint,aps,epsf,fixes]{revtex}
\begin {document}

\def\md{m_{\rm d}}
\def\CA{C_{\rm A}}
\def\p{{\bf p}}
\def\k{{\bf k}}
\def\x{{\bf x}}
\def\y{{\bf y}}
\def\z{{\bf z}}
\def\v{{\bf v}}
\def\n{{\bf n}}
\def\A{{\bf A}}
\def\E{{\bf E}}
\def\B{{\bf B}}
\def\D{{\bf D}}
\def\dn{{d n_0 \over d\Omega}}
\def\tr{{\rm tr}}
\def\real{{\rm real}}
\def\im{{\rm Im}\,}

\def\Omegap{\Omega_\p}
\def\lat{{\rm lat}}
\def\M{{\cal M}}
\def\nf{n_{\rm f}}
\def\ew{{\rm ew}}
\def\EulerGamma{\Gamma}
\def\soft{{\rm soft}}
\def\hard{{\rm hard}}
\def\gammamax{\gamma_{\rm max}}
\def\gammamin{\gamma_{\rm min}}

\preprint {UW/PT-97-2}

\title  {Hot B violation, the lattice, and hard thermal loops}

\author {Peter Arnold}

\address
    {%
    Department of Physics,
    University of Washington,
    Seattle, Washington 98195
    }%
\date {January 1997}

\maketitle
\vskip -20pt

\begin {abstract}%
{%
{%
It has recently been argued that the rate per unit volume
of baryon number violation (topological transitions) in
the hot, symmetric phase of electroweak theory is of the form
$\eta \, \alpha_{\rm w}^5 T^4$ in the weak-coupling limit, where
$\eta$ is a non-perturbative numerical coefficient.
Over the past several years, there have been attempts to extract the
rate of baryon number violation from real-time simulations of
classical thermal field theory on a spatial lattice.
Unfortunately, the coefficient $\eta$
will not be the same for classical lattice theories and the
real quantum theory.
However, by analyzing the appropriate effective theory on the lattice
using the method of hard thermal loops, I show that the {\it only} obstruction
to precisely relating the rates in the real and lattice theories is
the fact that the long-distance physics on the lattice is not
rotationally invariant.  (This is unlike Euclidean-time
measurements, where rotational invariance is always recovered in the
continuum limit.)  I then propose how this violation of rotational invariance
can be eliminated---and the real B violation rate measured---by
choosing an appropriate lattice Hamiltonian.
I also propose a rough measure of the systematic error to be expected
from using simpler, unimproved Hamiltonians.
As a byproduct of my investigation, the plasma frequency and
Debye mass are computed for classical thermal field theory on the lattice.
}%
}%
\end {abstract}


\section {Introduction}

   The violation of baryon number (B) in the hot, symmetric phase%
\footnote{
   Here and throughout, I use the term ``symmetric phase'' loosely since,
   depending on the details of the Higgs sector, there many not be any
   sharp transition between the symmetric and ``symmetry-broken'' phases
   of the theory \cite{kajantie,elitzur}.  A sharp transition is in fact
   required for electroweak baryogenesis.
}
of electroweak
theory plays a crucial role in scenarios for electroweak baryogenesis.
The rate of B violation is tied, through the electroweak anomaly, to the
the rate of topological transitions of the electroweak gauge fields.
For the symmetric phase, this rate is
not calculable by any perturbative method.
In this paper, I address whether the topological transition rate
can, in principle, be extracted from lattice simulations.
The discussion will reveal some surprising features of real-time
thermal field theory on the lattice.

It has long been appreciated that, at finite temperature, topological
transitions in real time are not directly related to topological transitions
in Euclidean time \cite{strike back}.
As a result, there is no apparent way to measure the
real-time thermal transition rate in a standard, Euclidean-time, lattice
simulation of quantum field theory.
Several years ago, Ambj{\o}rn and Krasnitz \cite{ambjorn}
cleverly implemented the observation \cite{grigoriev}
that a full
simulation of {\it quantum} field theory is not actually required.
Topological transitions occur in the symmetric phase through large
configurations which are essentially classical.  Indeed, {\it all}
long-distance bosonic physics in a hot, ultrarelativistic plasma
is effectively classical because of Bose enhancement.  The number
of quanta per mode in low-energy modes is given by the
Bose distribution
\begin {equation}
   n(E) \equiv {1\over e^{\beta E}-1} \sim {1\over\beta E} \gg 1
   \qquad \hbox{for} \qquad E \ll T
   \,,
\label {eq:classical regime}
\end {equation}
and, by the correspondence principle, this is the classical regime.
Unlike quantum field theory,
real-time simulations of classical field theory are tractable: just
evolve the classical equations of motion.

Classical thermal field theory has an ultraviolet catastrophe,
historically famous in the context of black-body radiation.
Because every mode has energy ${1\over2} T$ by the classical
equipartition theorem, and because there are an infinite number of
modes per unit volume in continuum field theory, the energy density
is infinite.  In quantum field theory, in contrast, the ultraviolet
contribution is cut off at momenta of order $T$.
Ambj{\o}rn and Krasnitz reasoned that the details of short-distance
physics should not affect the long-distance physics of topological
transitions.  In their simulations, they put their classical
system on a spatial
lattice and then evolved it in continuous real time.
The ultraviolet catastrophe was cut off by the lattice spacing,
which they progressively made smaller and smaller.

Son, Yaffe, and I \cite{alpha5}
have recently pointed out that short-distance effects
do not decouple as cleanly as Ambj{\o}rn and Krasnitz hoped.
In particular, the short-distance modes cause damping of the
long-distance dynamics, and this damping affects the transition
rate.  We showed that damping reduces the transition rate by a
factor of $O(\alpha)$, where $\alpha$ is the weak fine structure
constant.  Because the damping is caused by short-distance physics,
it is not universal, and so a comparison of theories with different
short-distance physics becomes non-trivial.
In the classical lattice theory used by Ambj{\o}rn and Krasnitz,
for example, damping reduces the rate instead by a factor of
$O(\alpha a T)$ where $a$ is the lattice spacing.
As I shall discuss, there is an even more serious problem: because
damping is dominated by short-distance physics,
it knows about the
anisotropies of the lattice.
As we shall see explicitly, and as was first noted by
Bodeker {\it et al.}\ \cite{bodeker}, the effective
long-distance physics of the classical lattice theory is not even
rotationally invariant.
This is in striking contrast to the familiar situation of Euclidean-time
simulations, where rotational invariance is always recovered in the
continuum limit.

I shall assume that our analysis of damping in ref.~\cite{alpha5}
is correct
and will eventually be borne out by numerical
simulations at sufficiently weak coupling.
In this paper, I focus on the natural follow-up question:
Given that the real quantum theory and the classical lattice theory
have different long-distance physics, is there any way to measure the
real topological transition rate?  The answer is yes---in principle---but
an exact calculation requires a careful choice of lattice action and
the numerical extraction of a somewhat awkward limit.%
\footnote{
  I shall briefly comment on the very different proposal of Hu
  and M{\"u}ller \cite{hu&muller} in my conclusion.
}
But I shall also propose a rough numerical measure of the suitability
of generic lattice actions and argue that
even results from simple, canonical actions
should be in the right ballpark if properly interpreted.

I shall not address at all the complicated problem of how one measures
topological transitions on the lattice in the first place, which has a long,
difficult history for Euclidean-time quantum simulations%
\footnote{
   For a brief review, see sec.~III.C.1 of ref.~\cite{schafer}.
   See also table I of ref.~\cite{vink}.
}
and a
shorter but still confusing one for real-time classical simulations
\cite{ambjorn,alpha5,Tang&Smit,Moore&Turok}.
I simply focus on dynamics and assume the measurement problem will eventually
be solved, using cooling or some other technique.

Section 2 of this paper outlines my method for taking a measurement of the
topological transition rate in a theory with one ultraviolet cut-off
({\it e.g.}\ a classical spatial lattice theory) and using it to predict the
rate in the same theory with a different ultraviolet cut-off ({\it e.g.}
the real, continuum, quantum field theory).  The procedure will
require, however, that the long-distance dynamics be rotationally
invariant in both cases, and this poses a problem for the lattice that
will eventually be dealt with.  Section 2 rests on a very rough and
schematic discussion of the effective long-distance dynamics,
and I return in section 3 to do a better job of reviewing the correct
long-distance theory.  I review the derivation of ``hard thermal loop''
effective theory but add a small twist.  The usual discussions in the
literature are based
on the assumption that hard ({\it i.e.}\ high-momentum)
excitations in the plasma
move at the speed of light.  This is not true for lattice theories,
and I show how the usual results easily generalize.
In section 4, as a warm-up example of a rotational-invariant classical
theory, I consider continuum classical thermal field theory with the
ultraviolet regulated by higher-derivative interactions.
In section 5, I turn to the canonical definition of the classical
theory on a simple cubic spatial lattice.
By explicit computation, the damping at long distances is shown to
be anisotropic and to have an interesting structure of
cusp and logarithmic singularities in its angular distribution.
I discuss the origin of these singularities.  Then I argue that
measurements of the topological transition rate in simple lattice theories
can still (if properly interpreted) be used to estimate the real transition
rate, and I propose a rough measure of the systematic error
arising from the anisotropy of the lattice.
In section 6, I propose that the real transition rate could in
principle be measured arbitrarily well from lattice simulations
by implementing a lattice version of the higher-derivative continuum
theory discussed in section 4.  In section 7, I briefly discuss
some possibilities of alternative lattice theories that may be
more rotational-invariant
than the canonical one, yet easier to implement than my proposal of
section 6.
Section 8 explores two quantities important to thermal field theory
but slightly tangential to the main thrust of this paper: the
plasma frequency and the Debye mass.  I compute both for classical,
thermal gauge theory on the lattice.  The results provide one test
of whether lattice simulations of a given size are indeed in the
small coupling limit, and I comment on the application of this test
to the simulations of ref.~\cite{ambjorn}.
Section 9 offers my conclusions and a summary of what
remains to be done.


\section {The Basics}

\label{sec:basics}

The origin of the basic scales associated with topological transitions
is reviewed in the introduction of ref.~\cite{alpha5}, and I shall
simply quote here the standard result: topological transitions proceed
through non-perturbative gauge configurations of spatial size
$R \sim 1/g^2 T$.  In ref.~\cite{alpha5}, we then showed that the
time scale associated with such transitions is $t \sim 1/g^4 T$
and that the transition occurs through slowly varying magnetic
configurations.%
\footnote{
  See also ref.~\cite{huet}.
}
(More technically, the relevant, low-frequency gauge fields are transversally
rather than longitudinally polarized.)
My current endeavor can be summarized as follows.  Consider
the effective long-distance and long-time theory corresponding to
frequencies and spatial momenta of order $(\omega,k) \sim (g^4T, g^2 T)$.
What is the correspondence of those theories between (a) the real
quantum theory, and (b) a cut-off classical theory?

The important characteristics of the long-distance effective theory
for topological transitions,
as discussed in refs.~\cite{alpha5,huet},
can be roughly introduced as follows:
The
behavior of the long-distance modes is analogous to a pendulum moving
in hot molasses, where the molasses represents the short-distance modes,
and non-perturbative physics such as the topological transition
rate corresponds to large-amplitude fluctuations of the pendulum.
A slightly non-linear pendulum is described by an equation of the form
\begin {equation}
   (\partial_t^2 + k^2) A + \hbox{(higher-order in $A$)} = 0\,.
\end {equation}
A pendulum in (idealized) hot molasses is described instead by a
Langevin equation of the form
\begin {equation}
   (\partial_t^2 + \gamma \partial_t + k^2 ) A
        + \hbox{(higher-order in $A$)} = \xi(t)\,.
\label{eq:pendulum}
\end {equation}
There is a damping term with coefficient $\gamma$ due to the viscous
interaction with the molasses.  There is also a random force
$\xi(t)$ due to the buffeting of the pendulum by random thermal
fluctuations of the molasses.  The two are related by the
fluctuation-dissipation theorem.  Ignoring the non-linearity of
the pendulum, the random force in (\ref{eq:pendulum}) must have
a white-noise spectrum with normalization $2\gamma T$:
\begin {equation}
   \langle \xi(t') \, \xi(t) \rangle = 2\gamma T \, \delta(t'-t) \,.
\label{eq:noise 0}
\end {equation}
The frequency response of such a pendulum is
\begin {equation}
   A(\omega) = {\xi(\omega) \over -\omega^2 - i\gamma\omega + k^2} \,,
\label{eq:response}
\end {equation}
with a power spectrum
\begin {equation}
   \langle A(\omega')^* \, A(\omega) \rangle
   = {2\gamma T \, \delta(\omega'-\omega) \over
       |-\omega^2 - i\gamma\omega + k^2|^2}
\end {equation}
which includes small-amplitude, high-frequency fluctuations and
larger-amplitude, low-frequency ones.
If the pendulum is strongly damped, $\gamma \gg k$, then the
characteristic inverse time for the largest amplitude fluctuations
is
\begin {equation}
   \omega \sim k^2/\gamma \ll k \,.
\label{eq:omega}
\end {equation}
In this case, the $\omega^2$ term is ignorable in (\ref{eq:response}),
and the large-amplitude fluctuations of the system can be described by
replacing (\ref{eq:pendulum}) with the simpler equation
\begin {equation}
   (\gamma \partial_t + k^2 ) A
        + \hbox{(higher-order in $A$)} = \xi(t) \,.
\label{eq:pendulum simplified}
\end {equation}

In the case of gauge theory, the damping coefficient%
\footnote{
   The reader should not confuse the damping coefficient $\gamma$
   with the plasmon damping rate: they refer to different kinematic
   regimes.  $\gamma$ describes the damping of
   nearly-static magnetic fields in the plasma, for which $\omega \ll k$
   and $\im\Pi$ is $O(g^2 T^2 \omega/k)$.
   In this case, $\im\Pi$ is dominated by
   hard contributions.
   The plasmon damping rate, on the other
   hand, describes the damping of propagating plasma waves, for which
   $\omega > k$ and $\im\Pi$ is $O(g^3 T^2)$
   \cite{braaten&pisarski}.
   In this case, the dominant contribution to $\im\Pi$ depends on soft
   as well as hard physics.
}
$\gamma$ is determined
from the imaginary part of the (retarded)
thermal self-energy $\Pi(\omega,{\bf k})$.
The effective equation of the long-distance modes is,
in frequency and momentum space,
\begin {equation}
   (-\omega^2 + \Pi(\omega,{\bf k}) + k^2 ) A
        + \hbox{(higher-order in $A$)} = \xi(\omega, {\bf k})\,,
\label{eq:schematic}
\end {equation}
where $k \equiv |\k|$.
The dominant contribution to $\Pi$ for $\omega,k \ll T$ is
well-known \cite{pi} and is generated by interactions with
short-distance (momenta $\sim T$) degrees of freedom.
In the frequency region of interest
($\omega \ll k \ll T$), and for transverse ({\it i.e.}\ magnetic)
fluctuations, it is given in weak coupling by
\begin {equation}
   \Pi(\omega,{\bf k}) \approx - i\omega \> {\pi\md^2 \over 4k} \,,
\label{eq:Pi}
\end {equation}
where $\md$ is the (leading order) Debye mass.
For the case of pure gauge theory,
\begin {equation}
   \md^2 = {\textstyle{1\over3}} \CA g^2 T^2 \,,
\label {eq:md 0}
\end {equation}
where $\CA$ is the adjoint Casimir and $\CA=N$ for SU($N$).
So, for each mode ${\bf k}$,
the damping coefficient $\gamma$ of (\ref{eq:pendulum})
is just
\begin {equation}
   \gamma = {\pi\md^2 \over 4k} \,.
\label{eq:gamma 0}
\end {equation}

For the spatial momentum scale of topological fluctuations,
$k = O(g^2 T)$,
(\ref{eq:omega}) confirms that $\omega \sim g^4 T \ll k$ and shows that
the effective theory can be described by a simplified Langevin
equation of the form (\ref{eq:pendulum simplified}).

So far, I have been very schematic.  Most significantly, I have blithely
ignored all the terms in the above equations labeled ``higher order in
$A$.''  Since I am ultimately interested in non-perturbative phenomena,
I need to do a better job.  I will review the full, unadulterated
equations of motion of the effective theory in the next section.
However, the schematic equations presented so far are adequate to
explain the basic idea of how to compare theories with different
ultraviolet behavior.

Ignoring the higher-order terms altogether
for now, consider two theories with different ultraviolet
cut-offs---say, the real quantum theory and some cut-off classical
theory.  The self-energy $\Pi$ will be different in the two
theories, and so the damping constant $\gamma$ will be different.
The long-distance effective equations of motion are then also
superficially different:%
\begin{mathletters}%
\label{eq:2 eqs}%
\begin {eqnarray}
   (\gamma_1 \partial_t + k^2 ) A &=& \xi_1(t)
\\
\noalign{\hbox{vs.}}
   (\gamma_2 \partial_t + k^2 ) A &=& \xi_2(t) \,.
\end {eqnarray}%
\end{mathletters}%
But, as I've written them, these two equations are trivially related by a
rescaling of time:
\begin {equation}
   t_1 \to {\gamma_1 \over \gamma_2} \, t_2 \,.
\label{eq:rescale}
\end {equation}
[That this transformation maps $\xi_1 \to \xi_2$ can be verified
from (\ref{eq:noise 0}).]  That means that the topological transition rates
$\Gamma$ of the two theories are related by a simple rescaling of time:
\begin {equation}
   \Gamma_1 = {\gamma_2 \over \gamma_1} \, \Gamma_2 \,.
\label{eq:convert 1}
\end {equation}
The prescription for measuring the real topological rate is then: (a) to
measure it in some UV cut-off classical theory, (b) to perturbatively compute
the self-energy and hence the values of $\gamma$ in both the real theory and
the classical theory, and then (c) to convert the measured rate by
(\ref{eq:convert 1}).

The fly in the ointment is that generically, in lattice theories,
$\gamma$ is not rotationally invariant but depends on the direction
of ${\bf k}$
with respect to the axis of the lattice.  I shall demonstrate
this later by explicit calculation.  This means that there is no
single rescaling of time that can relate the equations (\ref{eq:2 eqs})
for all the long-distance modes of the theories.  So there is another
step required in the prescription: (step 0) find a classical lattice
theory where the ultraviolet cut-off is as rotationally invariant as
possible.
I should emphasize that, unlike the more familiar case of Euclidean-time
simulations, this last step is a theoretical necessity and not
merely a numerical convenience.  In Euclidean simulations, improving
the rotational invariance of the action improves the rate of approach
to the continuum limit as the lattice spacing is decreased, but (in
principle) any lattice action will do as long as the lattice spacing is
small enough.  Here, that is not the case.

The chink in the armor of rotation invariance can already be seen in
the result (\ref{eq:Pi}) for the self-energy in the real,
rotational-invariant theory.  In Euclidean time, consider
effective interactions of the long-distance degrees of freedom that are
induced by the short-distance degrees of freedom.
Such interactions can be Taylor expanded
in the small, long-distance momenta $k_x$, $k_y$, and $k_z$.
The cubic symmetry of the lattice is enough to guarantee that any
interaction up to two derivatives $\k$ is in fact rotationally
invariant as well.  Interactions involving four derivatives need not be
rotationally invariant, but such interactions are irrelevant and decouple
from the long-distance physics.  The crucial assumption of this
argument is the analyticity of the interactions at $\k = 0$.
It has long been known that real-time thermal interactions generated
by short-distance physics {\it fail}
this analyticity criterion, as can be seen explicitly in (\ref{eq:Pi}).
The standard mechanism by which rotational invariance is recovered in
Euclidean time is therefore no longer operative.

Before proceeding to flesh out the proposals made in this
section, I should fix a point of nomenclature.
Often in this paper I shall refer to the ``real'' rate $\Gamma_\real$
of topological transitions in continuum, quantum, pure non-Abelian gauge theory
at high temperature and weak coupling.
If one's interest is in actual electroweak theory in the same limit,
with its accompanying Higgs and fermion fields, it is easy to convert
if one ignores the small effects of the Weinberg mixing angle.
The Debye mass in electroweak theory is \cite{carrington}
\begin {equation}
   m_{\rm d}^2 =
     \left( {\textstyle{5\over6}} + {\textstyle{1\over3}} \nf \right)
     g^2 T^2
\end {equation}
for a single doublet Higgs and $\nf=3$ families.  By the conversion
procedure (\ref{eq:convert 1}) outlined above and by (\ref{eq:gamma 0}),
the electroweak rate $\Gamma_\ew$ is related to the pure SU(2) gauge
theory rate ``$\Gamma_\real$'' by
\begin {equation}
   \Gamma_\ew = {\Gamma_\real \over
        \left( {\textstyle{5\over4}} + {\textstyle{1\over2}} \nf \right) } \,.
\label{eq:ew}
\end {equation}
Eq.~(\ref{eq:ew}) and the analysis of this paper
applies whenever the temperature is sufficiently high that
the infrared dynamics of the Higgs is irrelevant at lengths
of $O(1/g^2 T)$.  This is the case either (1) far above the electroweak
phase transition or crossover, or (2) in the symmetric phase at
the transition in cases where there is a 1st-order transition and
the transition is not exceedingly weak.  Other cases could be handled
by including the infrared dynamics of the Higgs into the discussion
and into simulations.


\section {Hard Thermal Loops}

\subsection {Review of Basic Derivation}
\label{sec:review}

It is now well-established how to write a set of non-local equations describing
the dominant interactions of long-distance modes in hot, real-time,
non-Abelian gauge theory when the random force term $\xi$ discussed
earlier is ignored.%
\footnote{
   The reader may wonder why numerical simulations don't simply
   simulate these equations directly.  The problem is their non-locality.
   For a discussion of the nightmare of trying to find a consistent,
   local description of the effective theory itself (which requires making
   precise the separation between soft and hard degrees of freedom), see
   ref.~\cite{bodeker}.  I shall be using the hard-thermal-loop
   equations only as
   a theoretical tool for comparing the (leading-order) long-distance
   dynamics of more fundamental, local theories.
}
These interactions go by the name of
``hard thermal loops,'' since they are generated by integrating out
the hard, short-distance modes of the theory.
Though there is a great deal of literature on this subject, it is
almost universally geared to hard particles that travel at the
speed of light.  In a lattice theory, in contrast, the hard particles have a
more complicated dispersion relationship than $\omega=k$.
It will therefore be useful to briefly review the derivation
of the hard thermal loop equations so that I can motivate the
(simple) modifications required for applying them to lattice theories.

Hard thermal loops were originally derived through diagrammatic analysis by
Braaten and Pisarski \cite{braaten&pisarski}.
For my purpose, the quickest way to get at them is
by an alternate method formalized by Blaizot and Iancu
\cite{blaizotQED,blaizotQCD}, which I shall
present heuristically.
(See also ref.~\cite{kelly}.)
The method is a generalization of the description
of QED plasmas via Vlasov equations.
The Vlasov equations are an effective description
where soft photons are represented by gauge fields $A_\mu(t,\x)$ in
the usual way, but all the hard, charged particles are represented by
a locally defined flow density $n(t,{\bf p},{\bf x})$, which is the
density of particles at time $t$ and position ${\bf x}$ with (hard) momentum
${\bf p}$.  The first Vlasov eq.\ follows from applying Liouville's Theorem
to $n$ and using the equations of motion:%
\footnote{
   For a discussion of why hard collision terms ({\it i.e.}\ terms non-linear
   in $n$) can be ignored in the application to topological transitions,
   see ref.~\cite{huet}.
}
\begin {eqnarray}
   0 ~=~ {d\over dt} \, n(t,\p,\x)
     ~=~ \partial_t n + \dot \x \cdot \partial_\x n
                    + \dot \p \cdot \partial_\p n
     ~=~ v^\mu \partial_\mu n
           + e ({\bf E} + {\bf v}\times{\bf B}) \cdot \partial_\p n \,,
\end {eqnarray}
where ${\bf v}$ is the 3-velocity and where $v^\mu \equiv (1,{\bf v})$.
(Note that $v^\mu$ does not transform as a Lorentz 4-vector.)
In the real world of ultrarelativistic plasmas, hard excitations move
at the speed of light and $\v$ is just $\hat\p = \p/|\p|$.  Here enters my
simple generalization of the standard discussion in the literature: for
more general underlying field theories, ${\bf v}$ should be identified
with the {\it group} velocity of the hard excitations, since it is the
group velocity which describes the physical rate of motion of charge.
If the dispersion relationship for the hard modes is $E = \Omegap$,
then the group velocity is
\begin {equation}
   {\bf v} = \nabla_\p \,  \Omegap \,.
\end {equation}
(For those who find this style of argument overly heuristic, an explicit
check of final results will be made in Appendix~\ref{apndx:diagrammatic}
using a diagrammatic approach.)

The second Vlasov equation, which is for the gauge field, is simply
\begin {equation}
   \partial_\mu F^{\mu\nu} = j^\nu = \sum_f e_f \int_\p v^\nu n_f \,,
\end {equation}
where I have added a flavor index $f$ to make explicit that there may
be different types of particles with different charges in the plasma.
I have also introduced the short-hand notation
\begin {equation}
   \int_\p \equiv \int {d^3 p \over (2\pi)^3} \,.
\end {equation}
Now linearize the Vlasov equations in the gauge fields and in deviations
$\delta n$ of $n$ from its equilibrium distribution $n_0$.  $n_0$
depends only on the energy $\Omegap$ of excitations and not on
$\x$ or $t$.  Using
\begin {equation}
   \partial_\p n_0 = \dn \, {\partial_\p \Omega}
                   = \dn \, \v \,,
\end {equation}
the linearized Vlasov equations are
\begin {mathletters}%
\begin {eqnarray}
   v \cdot \partial \, \delta n_f
           + e_f \, {\bf E} \cdot \v \, \dn = 0 \,,
\label{eq:linearized n}
\\
   \partial_\mu F^{\mu\nu} = \sum_f e_f \int_\p v^\nu \, \delta n_f \,.
\label{eq:linearized A}
\end {eqnarray}%
\end{mathletters}%
These equations can be solved formally for $\delta n$, yielding
\begin {equation}
   \partial_\mu F^{\mu\nu}(t,\x) = - \sum_f e_f^2
       \int_\p \dn \, {v^\nu v^j \over (v \cdot \partial + \epsilon)}
       \, E^j(t,\x) \,.
\end {equation}
$\epsilon$ is an infinitesimal that has been inserted to pick out the retarded
solutions to the equations.
For QED with a single massless fermion,
\begin {equation}
   \partial_\mu F^{\mu\nu} = - 4 e^2
       \int_\p \dn \, {v^\nu v^j \over (v \cdot \partial + \epsilon)}
       \, E^j \,,
\end {equation}
where the factor of $4$ counts the two polarizations each of the fermion and
anti-fermion.

It is easy to guess how the forgoing generalizes to non-Abelian gauge theories.
From (\ref{eq:linearized A}) it is clear that the $\delta n$ of interest must
carry an adjoint color index.  (Rather than a simple color-diagonal number
operator, $\delta n$ now represents a correlation between the colors that the
gauge bosons convert between.)  But then
the convective derivative $v \cdot \partial$ in
(\ref{eq:linearized n}) must be generalized to a
covariant convective derivative $v \cdot D$, where $D$ is understood
to act in the adjoint representation.  Similarly, the derivative in
(\ref{eq:linearized A}) should be covariant.  The resulting equation
for the soft modes is
\begin {equation}
   D_\mu F^{\mu\nu} = - 2 \CA g^2
       \int_\p \dn \, {v^\nu v^j \over (v \cdot D + \epsilon)} \, E^j \,,
\label{eq:blaizot}
\end {equation}
where $e^2$ has been replaced by $\CA g^2$ and the factor of 2 counts
the two helicity states of the hard gauge bosons.
Readers desiring a much more detailed justification should refer
to ref.~\cite{blaizotQCD}.

For our current application to topological transitions, where
frequencies are small compared to spatial momenta and the fluctuations
are magnetic, we are only interested in the small frequency limit
of (\ref{eq:blaizot}).  Working in $A_0 = 0$ gauge, this limit gives
\begin {equation}
   (\D \times \B)^i  =  2 \CA g^2
       \int_\p \dn \, {v^i v^j \over (\v \cdot \D + \epsilon)} \,
       \partial_t A^j \,.
\end {equation}
where $\B = \D \times \A$.
As long as the energy $\Omegap$ respects parity, the integral over
$\p$ would vanish if not for the $\epsilon$ factor.
Using
\begin {equation}
   {1\over \v\cdot\D + \epsilon}
    = P\, {1\over\v\cdot\D} + \pi\delta(\v\cdot i\D) \,,
\end {equation}
where $P$ stands for principal part,
the effective equation for the frequencies and momenta of interest
may be rewritten as
\begin {equation}
   \gamma^{ij}(i\D) \; \partial_t A^j + (\D\times\B)^i = 0 \,,
\label{eq:basic 1}
\end {equation}
where
\begin {equation}
   \gamma^{ij}(\k) \equiv - 2 \pi \CA g^2
       \int_\p \dn \, v^i v^j \, \delta(\v \cdot \k) \,.
\label{eq:gamma}
\end {equation}
This is indeed of the form of the
``non-linear system in molasses'' (\ref{eq:pendulum})
discussed schematically in section \ref{sec:basics},
except that it is missing the random force term.

\begin {figure}
\vbox
   {%
   \begin {center}
      \leavevmode
      
      \epsfbox [150 260 500 530] {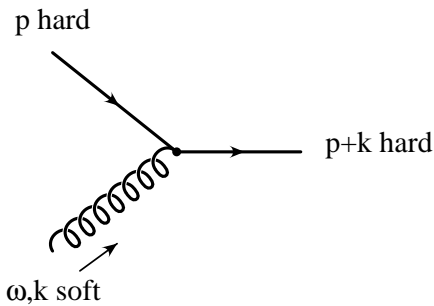}
   \end {center}
   \caption
       {%
         The absorption of a soft excitation (wavy line) by the small-angle
         scattering of a hard excitation (straight line).
       \label{fig:absorb}
       }%
   }%
\end {figure}

The underlying diagrammatic origin of the result (\ref{eq:gamma})
for the damping factor can be roughly understood by considering
the process of fig.~\ref{fig:absorb}, which shows a soft excitation being
absorbed by the small-angle scattering of a hard excitation.
The factor $g \, v^i$ in (\ref{eq:gamma}) comes from the interaction
vertex, and the second factor $g \, v^j$ is because the amplitude of
fig.~\ref{fig:absorb} should be squared to get a rate.
The $\delta(\v\cdot\k)$ is just the $\omega \ll k \ll p$ limit
of the energy-conserving delta function
$\delta(\Omega_{\p+\k} - \Omega_\p - \omega)$
associated with this diagram.
Finally, $dn/d\Omega$ for a Bose or Fermi distribution is just
$-\beta$ times the product
$n_0 (1 \pm n_0)$ of the initial-state probability distribution
and the final-state Bose enhancement or Fermi blocking factor.
More details are given in Appendix~\ref{apndx:diagrammatic}.

\subsection {The random force term}

The final result (\ref{eq:basic 1}) of the previous section is in fact
sufficient for the comparison of different theories, but, for the sake
of completeness, I will review the random force term as well.
The Vlasov equations of the previous section describe the relaxation or
propagation of deviations from equilibrium; they do not, as formulated
above, describe the fluctuations that occur in equilibrium itself.
However, fluctuations are related to dissipation through the
fluctuation-dissipation theorem.  Introduce a random force term by
writing
\begin {equation}
   \gamma^{ij}(i\D) \; \partial_t A^j + (\D\times\B)^i = \xi^i \,.
\label{eq:basic}
\end {equation}
If we momentarily ignore the
non-linearity of (\ref{eq:basic}), which arises from the non-Abelian
nature of the theory, then the characteristics of the
random force can be taken from (\ref{eq:noise 0}):
\begin {equation}
   \left\langle \xi^{ia}(t,\x) \, \xi^{jb}(t',\x') \right\rangle
    = 2 \, \delta^{ab} \, \gamma^{ij}\!(i \partial_\x) \, T \, 
      \delta^{(3)}(\x-\x') \, \delta(t-t') \,,
\end {equation}
where $a$ and $b$ are adjoint color indices.
Huet and Son \cite{huet} have shown how to modify this equation to
include the non-linearity and restore gauge invariance.
The result is to simply replace $\delta^{ab}$ above by a straight,
adjoint-charge Wilson line connecting $\x$ to $\x'$:
\begin {equation}
   \left\langle \xi^{ia}(t,\x) \, \xi^{jb}(t',\x') \right\rangle
    = 2 \, U^{ab}(t;\x,\x') \, \gamma^{ij}(\partial_\x) \, T \,  
      \delta^{(3)}(\x-\x') \, \delta(t-t') \,,
\label{eq:noise}
\end {equation}
where
\begin {equation}
   U(t;\x,\x') = {\cal P} \exp\left(
          - \int_\x^{\x'} d\z \cdot \A^a(t,\z) {\cal T}^a
          \right) \,,
\end {equation}
$\z$ runs on a straight path from $\x$ to $\x'$, ${\cal P}$ denotes path
ordering, and the ${\cal T}^a$ are the (real, anti-Hermitian) adjoint
representation generators.
The basic equations for the effective theory are then (\ref{eq:basic})
and (\ref{eq:noise}).


\subsection {Application to rotational-invariant theories}
\label{sec:app rot}

The damping coefficient $\gamma$ of (\ref{eq:gamma}) can be rewritten as
\begin {equation}
   \gamma^{ij}(\k) = {\pi M^{ij}(\hat \k) \over 4 |\k|} \,,
\end {equation}
where
\begin {equation}
   M^{ij}(\hat\k) \equiv - 8 \CA g^2
       \int_\p \dn \, v^i v^j \, \delta(\v \cdot \hat\k)
\label{eq:M}
\end {equation}
depends only on the direction $\hat\k$.
Note that $M$ is transverse:
\begin {equation}
   \hat k^i \,  M^{ij}(\hat\k) = 0 \,.
\end {equation}
In a theory that is fundamentally rotationally invariant, it must
then have the form
\begin {equation}
   M^{ij}(\hat k) = (\delta^{ij} - \hat k^i \hat k^j) \, m^2 \,,
\end {equation}
where $m^2$ is a number independent of $\k$, obtained
by averaging $\tr M$ over $\hat\k$:\thinspace%
\footnote{
  I am using a notation and normalization here that makes $m$ the analog
  of the $\md$ in (\ref{eq:Pi}).  However, $m$ is not actually the
  Debye mass in theories where $|\v| \not= 1$.
  See section \ref{sec:plasma freq}.
}
\begin {equation}
   m^2 = - 2 \CA g^2 \int_\p \dn \, |\v| \,.
\end {equation}
By examining the effective equations (\ref{eq:basic}) and
(\ref{eq:noise}), it is then easy to check that the time
rescaling suggested in section~\ref{sec:basics}
will indeed relate any two rotational-invariant theories:
\begin {equation}
   t_1 = {m_1^2 \over m_2^2} \, t_2 \,,
   \qquad
   \Gamma_1 = {m_2^2 \over m_1^2} \, \Gamma_2 \,.
\label{eq:convert}
\end {equation}

If $\Omegap$ is monotonically increasing, $m^2$ can be rewritten in the
simpler form
\begin {equation}
     m^2 = {2 \CA g^2 \over \pi^2} \int_0^\infty dp \> p \, n_0(\Omega_p) \,.
\label{eq:m}
\end {equation}
The result (\ref{eq:md 0}) for the real quantum theory can now be
reproduced by using the massless Bose distribution $(e^{\beta p}-1)^{-1}$
for $n_0(\Omega_p)$.


\section{An example: continuum theory with higher derivatives}
\label {sec:higher derivatives}

Before I move on to lattice theories and the difficulties with rotational
invariance, it is useful to first make the preceding discussion concrete with
a simple, rotational-invariant example of a classical thermal field theory.
Consider a classical continuum description of non-Abelian gauge
theory where I regulate the ultraviolet by higher-derivative terms.%
\footnote{
  The term ``higher derivative theory'' often refers to
  Lorentz-invariant theories, where the action must contain higher
  time derivatives as well as higher spatial ones.  Such theories create
  annoying issues about initial conditions and negative-norm states.
  The theory here, however, does not involve higher time derivatives and
  is free from such complications.
}
The Hamiltonian density is
\begin {equation}
   H = \tr \left[ E^i E^i
         + B^i \, f\!\left(\D^2\over\Lambda^2\right) \, B^i \right] \,,
\label {eq:high d}
\end {equation}%
where $f$ has an expansion
\begin {equation}
   f(z) = 1 + a_1 z + a_2 z^2 + \cdots
\end {equation}
and $\E = - \partial_t \A$ is (minus) the conjugate momentum to $\A$.
At momenta low compared to $\Lambda$, this is just normal gauge theory.
However, provided $f(z)$ grows sufficiently rapidly at large $z$, the
higher-derivative interactions will cut off the ultraviolet catastrophe of
classical thermal field theory at a momentum scale of order $\Lambda$.

To compare this cut-off classical theory to the real quantum theory,
I will compute the damping coefficient $m^2$ of (\ref{eq:m}).
For any classical theory, the equilibrium distribution $n_0(\Omega)$ is simply
\begin {equation}
   n_0(\Omega) = {T \over \Omega} \,.
\label{eq:n0 classical}
\end {equation}
One quick way to see this is to start with the Bose distribution of the
quantum theory and, for the first and last time in this paper, put in
the explicit factor of $\hbar$:
\begin {equation}
   n_0(\Omega) = {1\over e^{\hbar\beta\Omega} - 1}
   \to {T \over \hbar\Omega}
   \qquad \hbox{as} \qquad
   \hbar\to 0 \,.
\end {equation}
This just recapitulates the discussion (\ref{eq:classical regime})
of the introduction.
The remaining element we need is the frequency $\Omega_\p$ as a function
of momentum $\p$ for perturbative excitations.  For the theory
(\ref{eq:high d}), it is
\begin {equation}
   \Omega_p^2 = p^2 \, f\!\left(p^2\over\Lambda^2\right) \,.
\end {equation}
The coefficient $m^2$ of (\ref{eq:m}) is then (for monotonically
increasing $f$)
\begin {equation}
   m^2 = {2 \CA g^2 T \Lambda \over \pi^2}
         \int_0^\infty {ds \over \sqrt{f(s^2)} } \,.
\end {equation}
This converges provided $f(z)$ grows faster than $z$ for
large $z$.  In other words, a four-derivative interaction is not enough
to cut off the ultraviolet, but six derivatives do the job.
By (\ref{eq:convert}) and (\ref{eq:md 0}), the topological transition
rate $\Gamma_{\rm h.d.}$ in this classical, higher-derivative
theory is related (in weak coupling)
to the rate $\Gamma_\real$ in the real (pure gauge) quantum theory by
\begin {equation}
   \Gamma_\real = \Gamma_{\rm h.d.} \>
      {6 \Lambda\over\pi^2 T} \int_0^\infty {ds \over \sqrt{f(s^2)} } \,.
\end {equation}

It will be useful later on to have the specific result for $f$ of
the form $f(z) = 1 + a z^2$, corresponding to an isolated six-derivative
cut-off.  In this case,
\begin {equation}
   \Gamma_\real = \Gamma_{\rm h.d.} \>
      {3 \, \EulerGamma^2\!\left(\textstyle{1\over4}\right) \, \Lambda \over
       2 \, \pi^{5/2} \, a^{1/4} \, T} \,,
\label{eq:hd6}
\end {equation}
where $\EulerGamma(z)$ is the usual Euler Gamma function.


\section {A simple cubic spatial lattice}
\label{sec:SC}

\subsection {Results for \lowercase{$\gamma^{ij}$}}

We now have almost everything we need to compute the damping coefficient
$\gamma^{ij}$, or equivalently $M^{ij}$, for lattice theories; we just need
the dispersion relationship $\Omegap$.  Let's start by studying the simple,
classical, Kogut-Susskind Hamiltonian used by Amb{\o}rn and Krasnitz
\cite{ambjorn} for simulations on a simple cubic (SC), spatial lattice.
The tree-level dispersion relationship for propagating excitations
on the lattice is
\begin {equation}
   \Omegap^2 = 4 a^{-2} \left[
           \sin^2\left(p_x a\over2\right) +
           \sin^2\left(p_y a\over2\right) +
           \sin^2\left(p_z a\over2\right)
         \right] \,,
\label{eq:SC omega}
\end {equation}
where $a$ is the lattice spacing.
For thermal field theory on the lattice,
the parameters $g^2$, $a$, and $T$ always appear in the combination
$g^2 a T$ for dimensionless quantities.%
\footnote{
   This is because $g^2$ and $a$
   can be scaled out of the Hamiltonian $H$ so that the distribution
   $\exp(-\beta H)$ becomes $\exp(-H / g^2 a T)$.
}
I shall henceforth work in lattice units by setting $a$ = $T$ = 1.
The conversion back to physical ones%
\footnote{
   Lattice papers such as \cite{ambjorn} typically refer to
   the quantity $\beta_{\rm L} \equiv 2 N/g^2_\lat$ 
   for SU($N$) rather than $g^2_\lat$.
}
is $g_\lat \to g^2 a T$, $p_\lat \to ap$, $\omega_\lat \to a\omega$,
and $t_\lat \to a^{-1} t$.

The group velocity $\nabla\Omega$ is
\begin {equation}
   v^i = {\sin p^i \over \Omegap} \,.
\label{eq:SC v}
\end {equation}
Plugging into (\ref{eq:gamma}) with the classical distribution
(\ref{eq:n0 classical}) then yields%
\footnote{
   A general formula for the hard contribution to $\Pi^{ij}$ in a
   simple cubic lattice theory
   was first written down for QED coupled to massless
   scalars in eq.\ (B11) of ref.~\cite{bodeker}.
   My (\ref{eq:SC gamma}) is that equation's $\omega \ll k$ limit,
   once the overall normalization
   in ref.~\cite{bodeker} is changed to be appropriate for pure,
   non-Abelian gauge theory.
}
\begin{mathletters}%
\label{eq:SC gamma}%
\begin {eqnarray}
   \gamma^{ij}(\k) &=& {\pi M^{ij}(\hat\k) \over 4 |\k|} \,,
\\
   M^{ij}(\hat\k) &=& 8 \CA g^2
     \int_{-\pi}^{+\pi} {d^3p\over(2\pi)^3} \>
     {\sin p_i \, \sin p_j \over \Omegap^3} \,
     \delta(\hat\k\cdot\sin\p) \,,
\label {eq:SC M}
\end {eqnarray}%
\end{mathletters}%
where the integrals are over the Brillouin zone
$|p_i| \le \pi$.

\begin {figure}
\setlength\unitlength{1 in}
\vbox
   {%
   \begin {center}
   \begin {picture}(5,2.5)(0,-0.4)
      \put(-1,0){
        \leavevmode
        
        \epsfbox {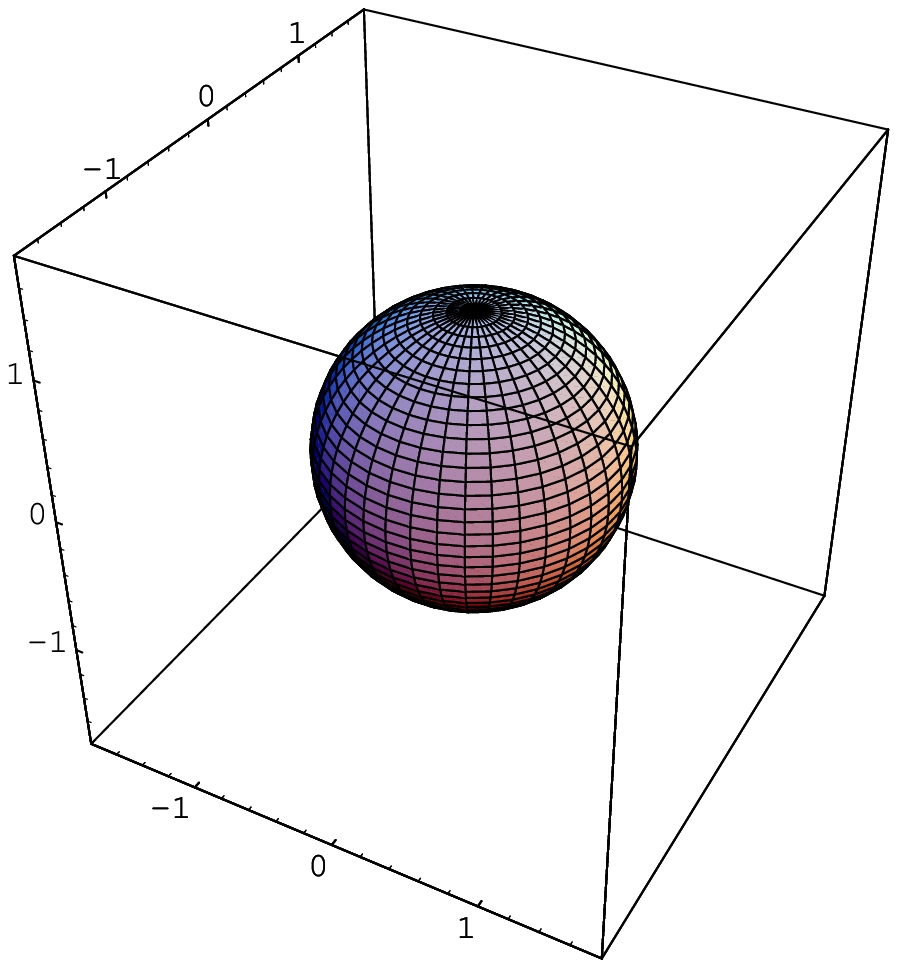}
      } 
      \put(1.4,0){
        \leavevmode
        
        \epsfbox {figimpib2.ps}
      }
      \put(3.8,0){
        \leavevmode
        
        \epsfbox {figimpic2.ps}
      }
      \put(0.0,-0.3){(a)}
      \put(2.4,-0.3){(b)}
      \put(4.8,-0.3){(c)}
   \end {picture}
   \end {center}
   \caption
       {%
       The trace $\gamma^{ii}$ of the damping coefficient as a function
       of direction, normalized by its angular average, for (a) a
       rotationally invariant theory, (b) a Kogut-Susskind
       Hamiltonian on an simple cubic lattice, and (c) for a face-centered
       cubic lattice.
       The spikes in (b) and (c) represent logarithmic divergences.
       \label{fig:impi}
       }%
   }%
\end {figure}

The delta function can be used to eliminate one of the three
integrations, and I have done the remaining two numerically.
(The detailed expression is given in Appendix~\ref{apndx:numerical}.)
Fig.~\ref{fig:impi}b shows the resulting angular distribution of the trace
$\gamma^{ii}$ normalized by its angular average
$\langle \gamma^{ii} \rangle$.
Not only is the result not rotationally invariant,
but it has an interesting structure of singularities as well.
There are cusps and, in certain special directions,
$\gamma^{ij}$ is infinite!  As we shall see, these singularities
are mild and integrable.  Also, they are true singularities only
in the weak-coupling limit: they are rounded off by higher-order
corrections but become sharper and sharper as $g \to 0$.


\subsection {The origin of singularities}
\label{sec:sing origin}

For any fixed direction $\hat\k$ of the soft momentum,
$\gamma^{ij}$ as given by (\ref{eq:gamma}) can be rewritten
\begin {equation}
   \gamma^{ij}(\k) \equiv - 2 \pi \CA g^2
       \int_S \dn \, {v^i v^j \over |\nabla_\p (\v \cdot \k)|} \,,
\label{eq:surface int}
\end {equation}
where the integral is over the surface $S$ in $\p$ space on which
$\v(\p)$ is orthogonal to $\k$.  If there is a point on this surface where
the vector $\nabla_\p (\v\cdot\k)$
vanishes, there will be a
singularity in $\gamma$.  If there is a entire line of points where
the vector vanishes, the singularity will be more severe.
This is analogous to Van Howe singularities in the density of states.%
\footnote{
  The density of states is $\rho(E) = \int_p \delta(E-\Omegap)$ and
  has singularities (typically cusps) when $\v = \nabla\Omegap$
  vanishes for some $\p$.
}
Naive counting of constraints suggests the existence of singularities
is to be expected:
5 degrees of freedom in $\p$ and $\hat\k$ versus 4 constraints in
$\v\cdot\hat\k = 0$ and $\nabla_\p (\v\cdot\hat\k) = 0$.

To summarize, singularities in $\gamma$ will be caused by momenta $\p$ such
that there is a null vector $\hat\k$, orthogonal to $\v = \nabla \Omegap$,
of the matrix $\nabla \v = \nabla \nabla \Omegap$.
In the case of the dispersion relationship (\ref{eq:SC omega}),
\begin {equation}
   \nabla^i \nabla^j \Omegap =
       { \delta^{ij} \cos p^i - v^i v^j \over \Omegap } \,.
\label{eq:SC matrix}
\end {equation}
At $\p = (\pi/2,\pi/2,\pi/2)$ and its lattice reflections, this matrix is
proportional to $v^i v^j$ and so every direction of $\hat\k$ orthogonal to
$\v \propto (1,1,1)$ is associated with a singularity of $\gamma$.
This is the origin of the cusp singularities in fig.~\ref{fig:impi}b,
which are
along the plane $k_x + k_y + k_z = 0$ and its three lattice reflections.

For $\p = (\pi/2,\pi/2,p_z)$ and general $p_z$, the matrix
(\ref{eq:SC matrix}) has (1,-1,0) as a null vector, which is orthogonal
to $v$.  Therefore, $\hat\k$ in the (1,-1,0) direction or its lattice
reflections is associated with an entire line of singularities in
$\p$ space.
This is the origin of the infinities in these $\hat\k$ directions in
fig.~\ref{fig:impi}b.
(See section \ref{sec:nospike} for a more general discussion.)
For these values of $\hat\k$,
$\nabla_\p(\v\cdot\hat\k)$ vanishes linearly as one approaches the
$\p$-space line of singularities.  The surface integral in
(\ref{eq:surface int}) is therefore only logarithmically singular.

Physically, there are not, in fact, infinite spikes in the angular
distribution of $\im\Pi(\omega,\k)$ in the lattice theory.  Recall that
$(\omega,k) \sim (g^4, g^2)$ for the long-distance physics of interest.  The
infinite size of the spikes occurs only in the formal limit
$\omega \ll k \ll 1$
that I applied to study the theory in weak coupling.  The taming of
the singularity can be seen from the Vlasov equations by returning to the
hard-thermal loop equation (\ref{eq:blaizot}) before extracting the small
$\omega$ limit.  The imaginary part of the self-energy, for spatial
polarizations, is
\begin {equation}
   \im\Pi^{ij} = 2 \CA g^2 \omega \int_\p \dn\, {v^i v^j} \delta(v \cdot K) \,,
\end {equation}
where $K$ is the four-vector $(\omega,\k)$.
The relevant difference from the earlier derivation of the damping
constant $\gamma^{ij}$ of (\ref{eq:gamma}) is simply that
$\delta(\v\cdot\k)$ has been replaced by
\hbox{$\delta(\v\cdot\k - \omega)$}.
This difference is sufficient to cut off the logarithmic infinities
in $\gamma^{ij}$ from order $g^2 k^{-1} \ln\infty$ to order
$g^2 k^{-1} \ln(k/\omega) \sim g^2 k^{-1} \ln(1/g^2)$.

It is beyond my present purpose to explore the cut off of the logarithmic
singularities in detail, but I should mention that the previous
paragraph is
misleadingly incomplete.  Based on the Vlasov equations, I've shown how
the logarithms are cut off by effects sub-leading in the limit
$\omega \ll k \ll 1$.  At sub-leading order, however, the Vlasov
equations themselves are inadequate.
One can see this from a more fundamental, diagrammatic analysis of
the self-energy, discussed in Appendix~\ref{apndx:diagrammatic}.
The $\delta(v\cdot K)$ discussed above is nothing more than
the $\omega,k \ll p$ limit of the energy-conserving delta
function $\delta(\Omega_{\p+\k} - \Omega_\p - \omega)$
associated with fig.~\ref{fig:absorb}
and discussed at the end of section~\ref{sec:review}.
For $(\omega, \k)$
of order $(g^4, g^2)$ in coupling, however,
the correct sub-leading version is
\begin {equation}
   \delta(\omega_{\p+\k} - \omega_\p - \omega)
   \approx
   \delta(\v\cdot\k - \omega + {\textstyle{1\over2}}\, \k\cdot\nabla\v\cdot\k)
   \,.
\end {equation}
The $\k\cdot\nabla\v\cdot\k$ term is the same order as the $\omega$
term.  Physically, it represents dispersion of the wave
packets of the hard excitations---an effect not built into the
Vlasov equations' implicit treatment of hard excitations as a
collection of classical particles.

In any case, the divergences in the small-coupling limit are only
logarithmic and so are mild enough not to cause any real problem
with the effective theory specified by (\ref{eq:SC M}).
Topological transitions occur through configurations with
spatial size $1/g^2 T$ and not through pure, infinite-extent, plane waves
with momenta $\k$ precisely in one of the divergent directions.
Since logarithmic singularities are integrable in $\hat\k$, they will give
only a finite damping effect to the evolution of any spatially localized
configuration.


\subsection {Estimating the real transition rate}

   If there were no infinite spikes, fig.~\ref{fig:impi}b
would look like at least a rough approximation to the
rotational-invariant sphere of fig.~\ref{fig:impi}a,
and a procedure for estimating the real, quantum transition
rate $\Gamma_\real$ suggests itself: replace $\gamma$ by its appropriate
angular average $\bar\gamma$, and use this value to relate
the lattice measurement of $\Gamma$
to $\Gamma_\real$ as in (\ref{eq:convert 1}) and
section~\ref{sec:app rot}.
What would be the expected systematic error of this procedure?
Suppose $\gammamax(k)$ and $\gammamin(k)$ were the maximum and the minimum
of the angular distribution of $\gamma(\k)$.
(I shan't worry about the indices $i,j$ on $\gamma^{ij}$ for the moment.)
Since the effect of damping decreases the transition rate, it seems
reasonable that the measured $\Gamma$ is bigger than it would have
been if $\gamma(\k)$ had been isotropically equal to $\gammamax$
in all directions but smaller than if $\gamma(\k)$ had been isotropically
equal to $\gammamin$ in all directions.  So the ratio of the
extreme values $\gammamin$ and $\gammamax$ to $\bar\gamma$ would
give a conservative estimate of the relative systematic error due
to the anisotropy of the damping.

   I shall discuss later whether one might find lattice theories
for which $\gamma(\k)$ indeed has no infinite spikes.
For the moment, focus on the case at hand.
Since the spikes are integrable, we can still replace $\gamma$
by its average to obtain an estimate of $\Gamma_\real$ from
the lattice measurement of $\Gamma$.
$\bar\gamma$ is the damping felt by isotropic gauge configurations.
The problem that remains
is to find a plausible estimate of the systematic error, so that
there is a measure for deciding whether one lattice theory with
spikes is better than another.  One obvious candidate is the
root-mean-square deviation $\sigma_\gamma$ given by
$\sigma_\gamma^2 = \overline{(\gamma - \overline{\gamma})^2}$,
where the overline indicates angular averaging.

   I have suppressed spatial indices on $\gamma$ above and will
now be more concrete.  By comparison with the rotational-invariant
case, where $\gamma^{ij}(\k)$ has the form
\begin {equation}
   \gamma^{ij}(\k) = (\delta^{ij} - \hat k^i \hat k^j) \, \gamma(k)
\end {equation}
due to its transversality, define
\begin {eqnarray}
   \bar\gamma^{ij}(\k) &\equiv&
          (\delta^{ij} - \hat k^i \hat k^j) \, \bar\gamma(k) \,,
\\
   \bar\gamma(k) &\equiv&
         {\textstyle{1\over2}} \langle \gamma^{ii}(\k) \rangle_{\hat\k} \,,
\end {eqnarray}
for the lattice case, where $\langle\cdots\rangle_{\hat\k}$ denotes
averaging over the direction of $\k$.  The variation is
\begin {equation}
   \sigma^2 \equiv
       {\textstyle{1\over2}}
       \langle \gamma^{ij}(\k) \, \gamma^{ij}(\k) \rangle_{\hat\k}
        -  [\bar\gamma(k)]^2 \,.
\end {equation}
Using the general result (\ref{eq:gamma}) for $\gamma^{ij}$ in any
theory, we have
\begin {equation}
   \bar\gamma(k) = - {\pi \CA g^2 \over 2k} \int_\p \dn \, |\v|
\label{eq:gamma bar}
\end {equation}
and
\begin {equation}
   {\sigma^2\over\bar\gamma^2} ~=~
      { \displaystyle
        \int_\p \int_{\p'} \dn \, {dn_0'\over d\Omega'}
           \, {4 (\v\cdot\v')^2 \over \pi|\v\times\v'|}
        \over \displaystyle
        \left(\int_\p \dn \, |\v|\right)^2
      }
      ~-~ 1
   \,.
\label {eq:sigma}
\end {equation}

For the Kogut-Susskind Hamiltonian on a simple cubic lattice,
numerical integration gives $\sigma/\bar\gamma = 0.31$ and
\begin {equation}
    \bar\gamma(k) = 0.2687 \, {\CA g^2 \over k} \,.
\end {equation}
Using (\ref{eq:convert}) and (\ref{eq:gamma 0}),
one can then estimate the real transition rate
(in the weak coupling limit) from a lattice measurement as follows.
The B violation rate $\Gamma$ is expected to scale as $\alpha^5$
in weak coupling \cite{alpha5}.
In lattice units, measure the proportionality constant
\begin {equation}
   \eta_1 \equiv \lim_{g_\lat\to 0} \, {\Gamma_\lat \over g_\lat^{10}} \,.
\end {equation}
Then the estimate is
\begin {equation}
   \Gamma_\real \approx {12\times 0.2687 \over \pi} \; \eta_1 g^{10} T^4 \,.
\label{eq:estimate}
\end {equation}
The fact that the rough indicator $\sigma/\bar\gamma$ of the systematic
error due to using the lattice is about $30\%$ suggests that
the estimate (\ref{eq:estimate}) should at least be in the right
ballpark of the true answer.
(Also, the minimum value of $\tr\,\gamma(\hat\k)$ in
fig.~\ref{fig:impi}b is just 30\% below the average.)
Unfortunately, $\eta_1$ has not yet been measured,%
\footnote{
   $\eta_1$ should not be confused with the coefficient $\kappa$ presented in
   ref.~\cite{ambjorn}, which was an attempt to extract the coefficient
   of a presumed $\Gamma \sim \alpha^4$ scaling law for weak
   coupling---a scaling law which ignores the effects of damping
   \cite{alpha5}.
}
and even the weak-coupling
scaling law $\Gamma \sim \alpha^5$ has yet to be verified on the
lattice.


\section {Achieving the rotational-invariant limit}
\label{sec:proposal}

To eliminate the fundamental systematic uncertainty caused by the anisotropy of
the lattice, one needs to find an effectively rotational-invariant
lattice theory.  (I re-emphasize that the search for an alternative
theory is a theoretical necessity: just taking the small lattice spacing
limit is by itself inadequate.)
As it turns out, an effectively rotational-invariant theory is,
in principle, easy to achieve.
The idea is simply to start with a rotational-invariant continuum
theory with a higher derivative cut-off, such as discussed in
section~\ref{sec:higher derivatives}, and then put it on a lattice
that is fine compared to the continuum cut-off scale
$\Lambda^{-1}$.
A numerical extraction of $\Gamma$ would then require
a careful double limit:
\begin {equation}
   \lim_{a\Lambda\to0} ~ \lim_{g_\lat\to0} ~ \Gamma_\lat \,,
\end {equation}
where the order of limits is crucial.

To be a little more specific about how the cut-off $\Lambda$ might be
achieved on a spatial lattice, consider the tree-level dispersion
relationship for propagating gluons with momentum small compared to
the lattice spacing.  For a given lattice Hamiltonian (with cubic
symmetry), it will have
an expansion in momentum of the form
\begin {equation}
   s \Omega_\p^2 = b |\p|^2
      + \left[ c_1 |\p|^4 + c_2 (p_x^4 + p_y^4 + p_z^4) \right]
      + O(p^6)
\end {equation}
in lattice units,
where the coefficient $s$ is just a reminder that the units of time
can be normalized arbitrarily.
By including next-to-nearest-neighbor couplings,
and perhaps next-to-next-to-nearest
neighbor couplings, and so forth, the parameters of the interactions can
be tuned to make any finite set of the coefficients $b,c_1,c_2,\cdots$
of the expansion take on whatever values desired.
In particular, there is some
local Hamiltonian $H_\epsilon^{(4)}$ for which (1) $b$ is a small number,
which I'll call $\epsilon^2$, (2) $c_1$ is $O(1)$, (3) $c_2 = 0$, and
(4) all other coefficients are $\le O(1)$:
\begin {equation}
   s \Omega_\p^2 = \epsilon^2 |\p|^2 + c_1 |\p|^4 + O(p^6) \,.
\label{eq:4 deriv}
\end {equation}
Choosing units of time where $s=\epsilon^2$,
\begin {equation}
   \Omegap^2 = |\p|^2
       + {c_1\over\epsilon^2} \, |\p|^4
       + O\left( {1\over\epsilon^2} p^6 \right) \,.
\label {eq:4 deriv 2}
\end {equation}
This describes a
theory with a rotational-invariant four-derivative ``cut off'' at momentum
$\Lambda = \epsilon a^{-1}$, plus anisotropic terms that are suppressed by
$\epsilon^2$ at that scale.

Unfortunately, the four-derivative ``cut-off'' in (\ref{eq:4 deriv 2}) is
inadequate.  As discussed in section~\ref{sec:higher derivatives},
a six-derivative interaction is required to cut off the UV contribution to
damping.  So, even for small $\epsilon$, a generic theory with dispersion
relationship (\ref{eq:4 deriv}) will still produce significant
anisotropy.%
\footnote{
   The relative contribution of $\p \sim 1$ to the
   damping rate is actually $O(1/\ln\epsilon)$ and so, technically, is small in
   the limit that $\ln\epsilon$ was very large.  But this is certainly
   an impractical limit for any numerical simulation, keeping in mind
   that, as discussed below, the $\epsilon \to 0$ and $g_\lat \to 0$ limits
   do not commute.
}
We can fix the cut-off by going to six derivatives, requiring improved
Hamiltonians $H_\epsilon^{(6)}$ where
(1) $b$ is a small number,
which I'll now call $\epsilon^4$,
(2) the fourth-order coefficients $c_1$ and $c_2$ vanish,
(3) the coefficient of $|\p|^6$ is $O(1)$,
(4) the sixth-order coefficients that break rotational invariance vanish,
and (5) other coefficients are $\le O(1)$:
\begin {equation}
   s \Omega_\p^2 = \epsilon^4 |\p|^2 + d_1 |\p|^6 + O(p^8) \,,
\label{eq:6 deriv}
\end {equation}
Choosing units of time where $s=\epsilon^4$,
\begin {equation}
   \Omegap^2 = |\p|^2
       + {d_1\over\epsilon^4} |\p|^6
       + O\left({1\over\epsilon^4} p^8\right) \,.
\label {eq:6 deriv 2}
\end {equation}
This describes
the desired rotational-invariant six-derivative cut-off at momentum
$\Lambda = \epsilon a^{-1}$ plus anisotropies that are suppressed by
$\epsilon^2$ at that scale.  The relative size of the anisotropies in
the damping coefficient $\gamma$ will therefore be $O(\epsilon^2)$ and
disappear in the limit $\epsilon \to 0$.

I will not attempt in the present work to explicitly construct the
improved lattice Hamiltonian $H_\epsilon^{(6)}$ which gives this
behavior.  A very nice discussion of improving lattice Hamiltonians
is given by Moore in ref.~\cite{moore improved}, where he constructs
a classical Yang-Mills Hamiltonian with vanishing $p^4$ terms.
Constructing a Hamiltonian that satisfies the other constraints
for $H_\epsilon^{(6)}$ is presumably straightforward in principle but
tedious in practice.

In everything above, I have referred to conditions on the
tree-level dispersion relationship.
The real dispersion relationship for hard excitations ($p \sim \Lambda$)
will be modified by perturbative corrections, whose strength is
parametrized by $g_\lat / \Lambda a = g_\lat / \epsilon$.
This is the reason that taking the $g_\lat \to 0$ limit
before the $\Lambda a \to 0$ limit would be crucial to extracting
results from numerical simulations.

For an improved Hamiltonian giving the desired low-momentum
dispersion relation (\ref{eq:6 deriv}), the conversion of
lattice measurements to the real transition rate would be achieved
by first measuring the coefficient
\begin {equation}
   \eta_2 \equiv \lim_{\epsilon\to0} \, \lim_{g_\lat\to 0} \,
       {\epsilon \, \Gamma_\lat \over g_\lat^{10}}
\end {equation}
and then taking
\begin {equation}
   \Gamma_\real =
      {3 \, \EulerGamma^2\!\left(\textstyle{1\over4}\right) \, \eta_2 \over
       2 \, \pi^{5/2} \, d_1^{1/4} }
         \; g^{10} T^4 \,,
\end {equation}
which follows from (\ref{eq:hd6}).
$g_\lat^2$ above continues to mean $g^2 a T$ where $g^2$ is the
continuum coupling extracted from the Hamiltonian $H_\epsilon^{(6)}$.


\section {Other lattice theories}

\subsection {The FCC lattice}

The $a\Lambda \to 0$ limit discussed in the last section may be
difficult to extract numerically.
So it's worthwhile to consider whether there are relatively simple
lattice theories where the damping coefficient $\gamma$, though
anisotropic, is closer to rotational invariance than for the
simple lattice Hamiltonian discussed in section~\ref{sec:SC}.
I will briefly discuss here one possibility for improvement:
replacing the simple cubic lattice by a face-centered cubic (FCC)
lattice.%
\footnote{
   In four-dimensional Euclidean gauge theories, people have tried
   different lattice types for the merely practical reason of trying to
   speed approach the continuum limit.  Lattices used include the
   F$_4$ lattice \cite{F4},
   which is the four-dimensional analog of an FCC lattice,
   and the body-centered hypercubic (BCH) lattice \cite{BCH}.
}
The FCC lattice can plausibly lead to more rotational-invariant results
because sites have more nearest neighbors than for the simple cubic
lattice (eight instead of six).

The formulation of lattice QED on a spatial FCC lattice is discussed
by Gosar in ref.~\cite{gosar}.  The fundamental plaquettes of an FCC lattice
are equilateral triangles, and Gosar studies a simple Kogut-Susskind
Hamiltonian defined on those plaquettes.
All I need here is the dispersion relation, which will be the same
for a non-Abelian gauge theory as for photons in QED and is given by Gosar.
For a simple cubic lattice, there are three orientations of links
($\hat x$, $\hat y$, and $\hat z$) and three branches of excitations:
two, degenerate, propagating polarizations with energy 
(\ref{eq:SC omega}); and one non-propagating polarization with
$\Omegap = 0$.  For an FCC lattice there are six orientations of links
and so six branches of excitations \cite{gosar}:
\begin {eqnarray}
   \hbox{(a)}   ~~~ \Omegap^2 &=& 0                     \,,
\\
   \hbox{(b)}   ~~~ \Omegap^2 &=& 32                    \,,
\\
   \hbox{(c,d)} ~~~ \Omegap^2 &=&
      32 \left[ 1 - \sqrt{1-\epsilon_\p/16} \right]     \,,
\label {eq:FCC Omega-}
\\
   \hbox{(e,f)} ~~~ \Omegap^2 &=&
      32 \left[ 1 + \sqrt{1-\epsilon_\p/16} \right]     \,,
\label {eq:FCC Omega+}
\end {eqnarray}
where $\epsilon_\p$ is the energy a massless scalar field with nearest-neighbor
coupling would have on an FCC lattice,
\begin {equation}
   \epsilon_\p \equiv 4 \left[ 3
     - \cos{p_x\over2} \, \cos{p_y\over2}
     - \cos{p_y\over2} \, \cos{p_z\over2}
     - \cos{p_z\over2} \, \cos{p_x\over2} \right] \,.
\label {eq:FCC epsilon}
\end {equation}
I have used lattice units where the side of a unit cell is $a = 1$.%
\footnote{
   Ref.~\cite{gosar} instead uses units where the distance between
   nearest neighbors is 1.
}
In the long wavelength limit,
branches (c,d) correspond to the two polarizations of the continuum
case $\Omega = |\p|$.  The non-propagating
branches (a) and (b) do not contribute to the damping coefficient $\gamma$
because their group velocity is zero.

The shape of the first Brillouin zone of an FCC lattice is slightly
complicated (a ``truncated octahedron''),
but all that is needed here is that the cube
$|p_i| \le 2\pi$ in momentum space covers the unique momentum states
exactly twice.  So, when integrating over states, I take
\begin {equation}
   \int_\p ~~\to~~ {1\over2} \sum_{\pm} \int_{-2\pi}^{+2\pi}
      {d^3p\over (2\pi)^3} \,.
\end {equation}
The sum is a notational reminder to sum over the two dispersion relationships
\begin {equation}
   \Omegap^2 =
      32 \left[ 1 \pm \sqrt{1-\epsilon_\p/16} \right] \,.
\label{eq:FCC omega}
\end {equation}
The corresponding group velocities are
\begin {equation}
   v_x = \pm \, {\displaystyle
              \sin{p_x\over2} \left( \cos{p_y\over2} + \cos{p_z\over2} \right)
              \over \Omegap \sqrt{1-\epsilon_\p/16} }
\label{eq:FCC v}
\end {equation}
and its permutations.

Numerical evaluation of the general formula (\ref{eq:gamma})
for $\gamma$
(see Appendix~\ref{apndx:numerical}) gives the angular distribution of
$\tr\,\gamma$ shown in fig.~\ref{fig:impi}c.
The position of the logarithmic spikes is explained
in the next section.
Numerical integration of (\ref{eq:gamma bar}) and (\ref{eq:sigma})
gives $\bar\gamma(k) = 0.501(1) \, \CA g^2 / k$ and\
$\sigma_\gamma/\bar\gamma = 0.26(1)$.
The 26\% result for the angular deviation of damping on the FCC lattice
is indeed better than the 31\% result for the SC lattice, but not
by a lot.  The minimum of $\tr\,\gamma(\hat\k)$ in
fig.~\ref{fig:impi}c is 20\% below the average.


\subsection {Avoiding logarithmic spikes}
\label{sec:nospike}

The simplest actions on both the SC and FCC lattices have logarithmic
spikes in the angular distribution of $\gamma$, shown in
figs.~\ref{fig:impi}b and c.
In the proposal of section~\ref{sec:proposal}
for reaching the rotational-invariant
limit, it doesn't matter whether the hardest modes $\p \sim 1$ of
the theory give rise to such spikes, because the effect of the
hardest modes, and hence the strength of the spikes, will vanish
as $\epsilon \to 0$.
For the purposes of surveying simpler lattice theories that are
only approximately rotationally invariant, however, it's interesting
to consider whether it's possible to avoid the spikes altogether.
So I will take a moment to explain the generic property of the
simple SC and FCC actions that causes the spikes and then discuss
whether it can be avoided.

\begin {figure}
\setlength\unitlength{1 in}
\vbox
   {%
   \begin {center}
   \begin {picture}(5,2.5)(0,-0.4)
      \put(0,0){
        \leavevmode
        
        \epsfbox {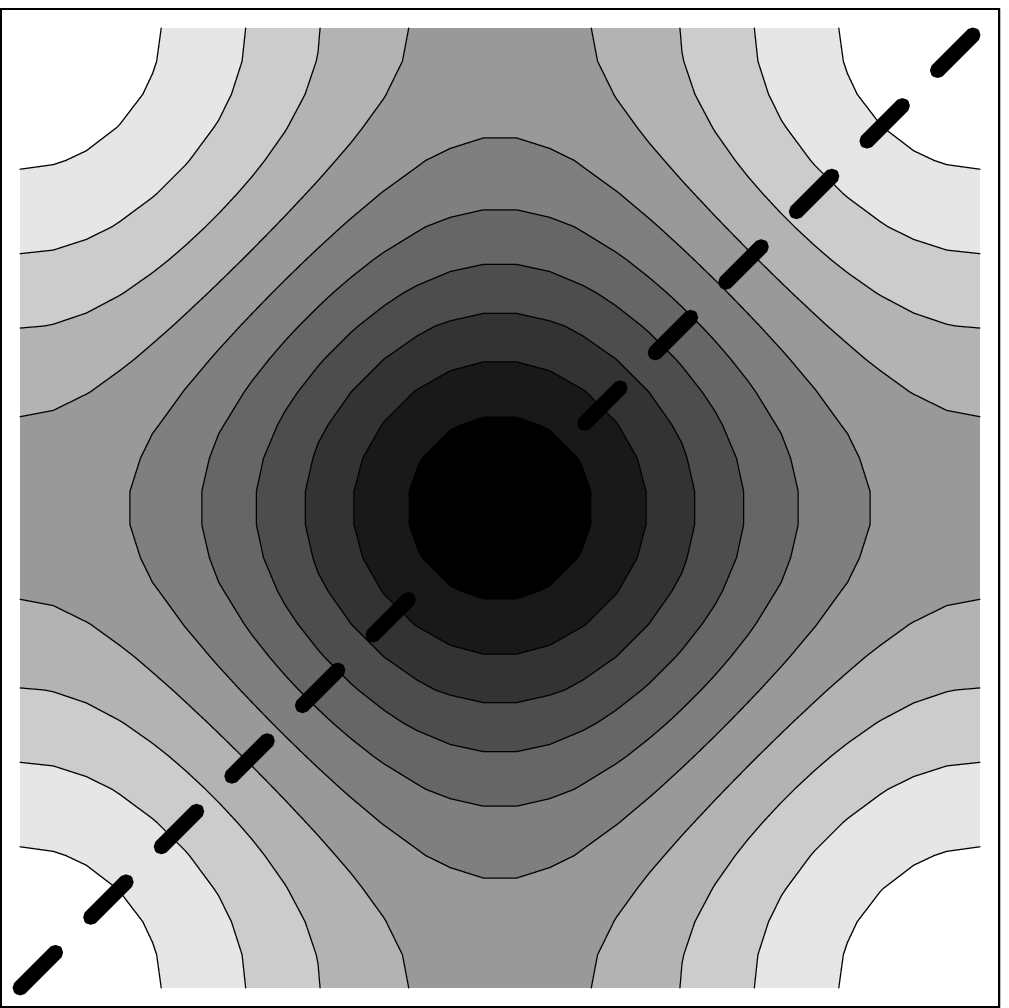}
      } 
      \put(2.8,0){
        \leavevmode
        
        \epsfbox {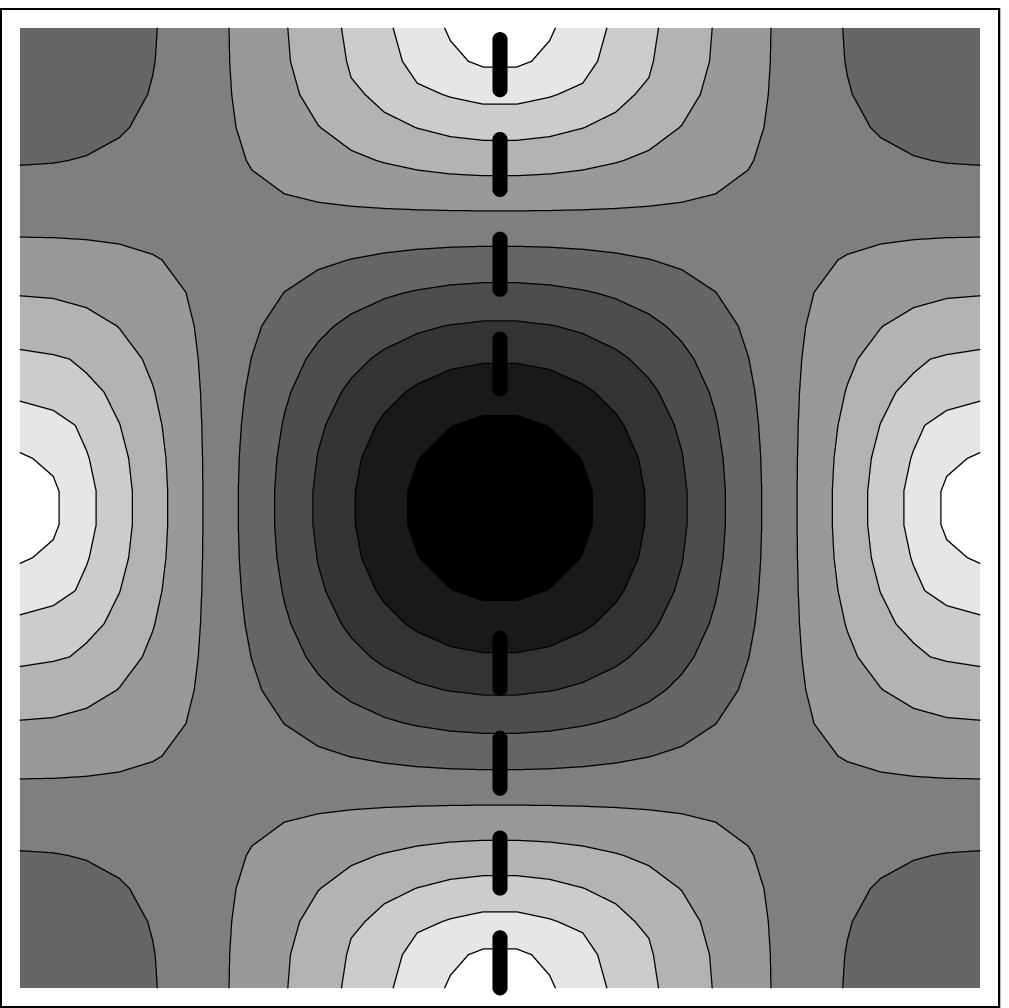}
      }
      \put(1.0,-0.3){(a)}
      \put(3.8,-0.3){(b)}
   \end {picture}
   \end {center}
   \caption
       {%
       A qualitative contour plot of $\Omegap$ vs. $p_x$ and $p_y$ for
       small, fixed $p_z$ in two different lattice theories.
       Black corresponds to low values of $\Omegap$ and white to high ones.
       \label{fig:spike}
       }%
   }%
\end {figure}

Recall the conditions $\hat\k \cdot \nabla\Omega = 0$ and
$\hat\k \cdot \nabla \nabla \Omega = 0$ of section~\ref{sec:sing origin} for
generating some sort of singularity.
Consider the energy $\Omegap$ of excitations for small but fixed
$p_z$ in a lattice theory with cubic symmetry.
Fig.~\ref{fig:spike}a shows a qualitative contour plot
of energy $\Omegap$ vs.\ $p_x$ and $p_y$ for some such lattice theory
[{\it e.g.}\ the simple cubic theory of (\ref{eq:SC omega})].
There is a minimum at $p_x = p_y = 0$, which there must be if
we are plotting the branch of $\Omegap$ that
approaches the continuum limit $\Omegap = |\p|$ for small $\p$.
Suppose that there is also a maximum somewhere along
the line $p_x = p_y$, as shown in fig.~\ref{fig:spike}a.
Then one can argue as follows that there must be a singularity associated with
the direction $\hat\k = (1,-1,0)$ and with $\p$ somewhere on the line
$p_x = p_y$ ($p_z$ still fixed).
By reflection symmetry through the plane $p_x = p_y$, the group
velocity $\nabla\Omega$ on the line must be perpendicular to $\hat\k$,
and the curvature $\nabla\nabla\Omega$ on the line must have the
block-diagonal form
\begin {equation}
   \nabla\nabla\Omega = \pmatrix{ A & B & 0 \cr
                                B & A & 0 \cr
                                0 & 0 & C } \,,
\end {equation}
in the basis $(p_z, p_x{+}p_y, \hat\k)$.  Near the minimum
in the fixed $p_z$ plane, $C$ must be negative.  Near the maximum that
by assumption lies on the line $p_x = p_y$, it must be positive.
Therefore, there is a point inbetween where $C=0$ and so
$\hat\k \cdot \nabla\nabla\Omega = 0$.  This point then satisfies both
conditions for generating a singularity associated with the
direction $\hat\k = (1,-1,0)$.

But now note that this argument did not depend on the value of $p_z$.
As long as the assumption that both a minimum and a maximum in
$(p_x,p_y)$ simultaneously lie
on the line $p_x = p_y$ holds for some {\it range} of $p_z$, then there
will be an entire curve of points in $\p$ space that contributes to the
singularity in the $\hat\k = (1,-1,0)$ direction.  As discussed earlier,
such degeneracy gives rise to a logarithmic singularity in $\gamma$.

A similar qualitative situation, for some different lattice theory
[{\it e.g.}\ the FCC theory of (\ref{eq:FCC Omega-})], is shown in
fig.~\ref{fig:spike}b.  The same argument now goes through with
the direction $\hat\k = (1,0,0)$ and the pane of symmetry $p_x = 0$.
This is the origin of the spikes in the FCC case of fig.~\ref{fig:impi}c.


To avoid spikes, look for lattice theories such that $\Omegap$
does {\it not} have lines of local maxima and minima in $(p_x,p_y)$ that
simultaneously lie on the same plane of symmetry.
To show that this is possible, I will give an example from massless scalar QED
on a simple cubic lattice (where momenta are $|\p_i| \le \pi$).
For scalar QED, the problem can be reduced to searching through the
space of $\Omegap^2$ that are finite linear combinations of
$\sin^2(\n\cdot\p/2)$ for integer vectors $\n$.
This is because one can generate any such $\Omegap^2$,
\begin {equation}
   \Omegap^2 = \sum_\n c_\n \sin^2\left(\n\cdot\p \over2\right) \,,
\end {equation}
by a gauged version of the local scalar interaction
\begin {equation}
   H_{\rm scalar} =
     \sum_\x |\Pi_\x|^2
     + \sum_\x \sum_\n {c_\n \over 4} |\phi_\x - \phi_{\x+\n}|^2 \,,
\end {equation}
where $\Pi$ is conjugate to $\phi$.
So I will therefore first construct an $\Omegap$ that works and
then leave it as an exercise to the interested reader to construct
the corresponding Hamiltonian.
The space of linear combinations of $\sin^2(\n\cdot\p/2)$
is equivalent, by trigonometric identities, to the space
of parity-even polynomials in $\cos p_i$ and $\sin p_i$.

\begin {figure}
\setlength\unitlength{1 in}
\vbox
   {%
   \begin {center}
   \begin {picture}(5,2.5)(0,-0.4)
      \put(-0.75,0){
        \leavevmode
        
        \epsfbox {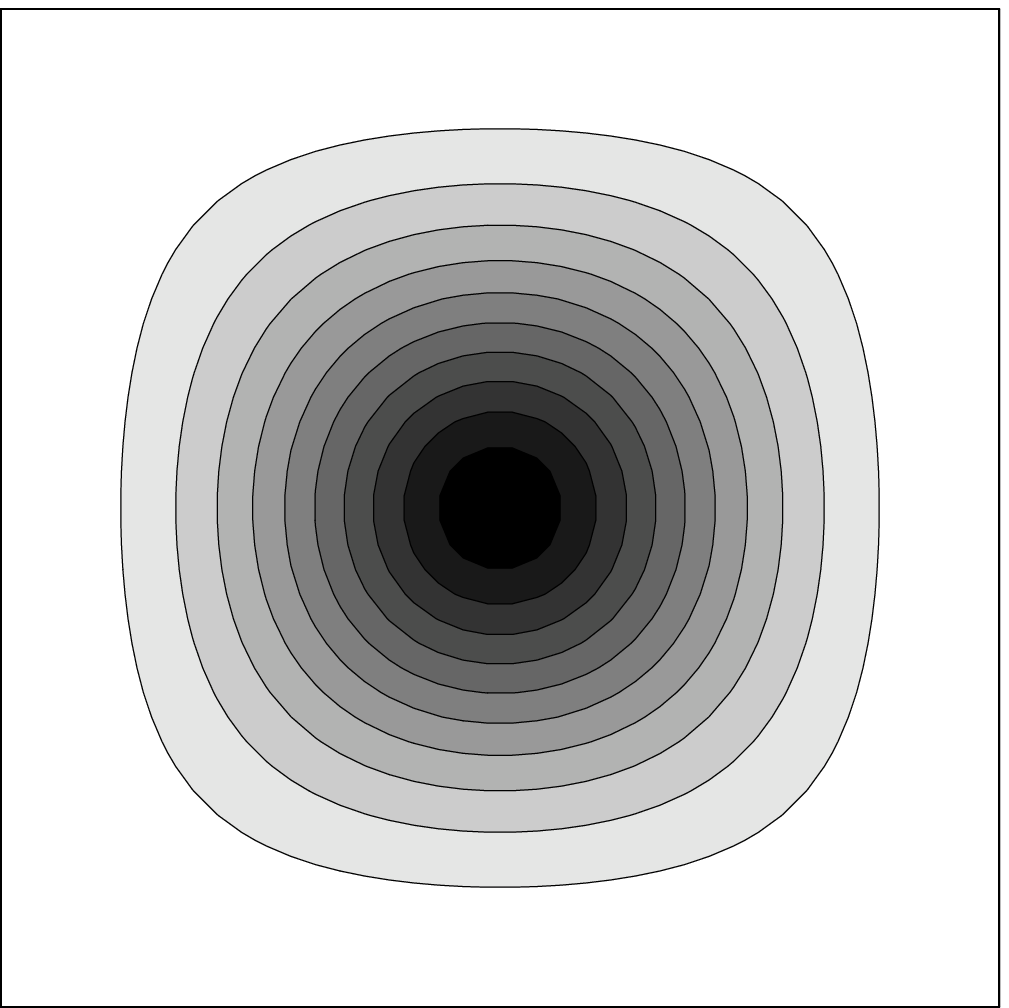}
      } 
      \put(1.45,0){
        \leavevmode
        
        \epsfbox {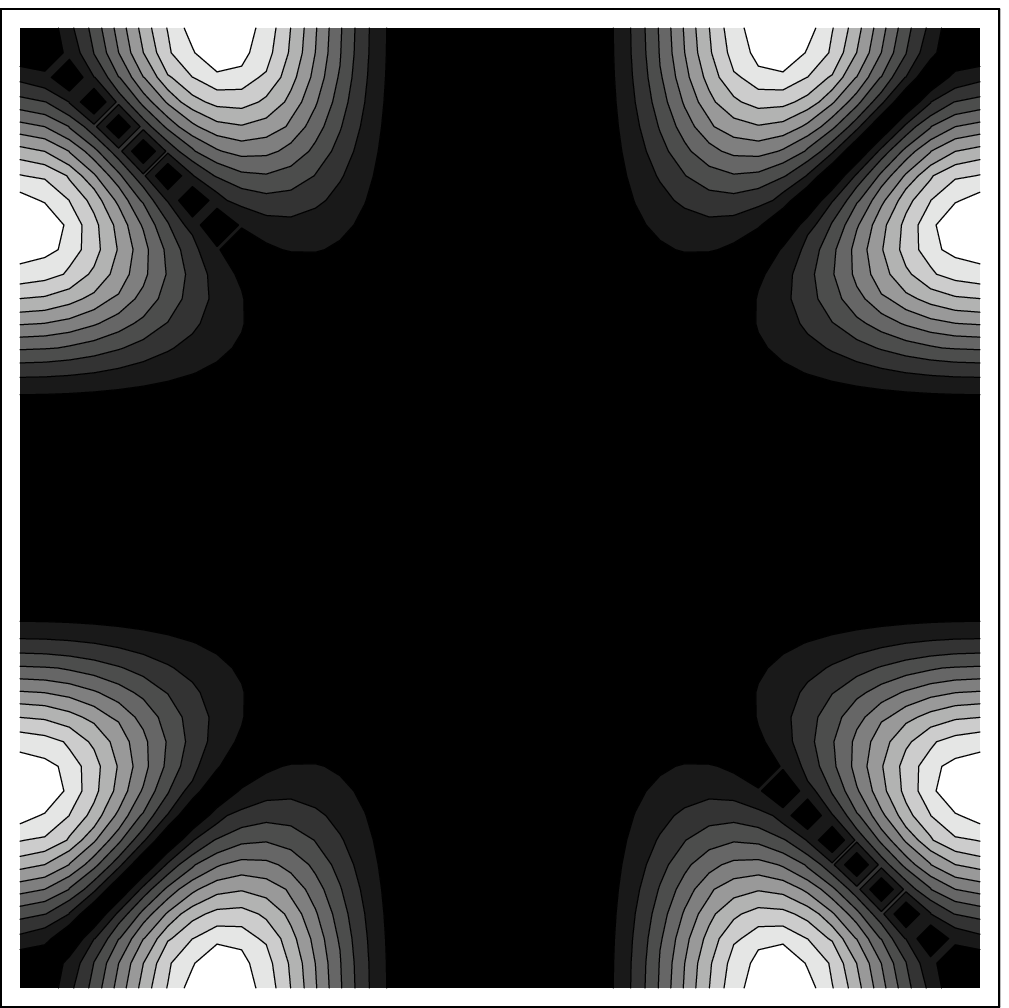}
      }
      \put(3.65,0){
        \leavevmode
        
        \epsfbox {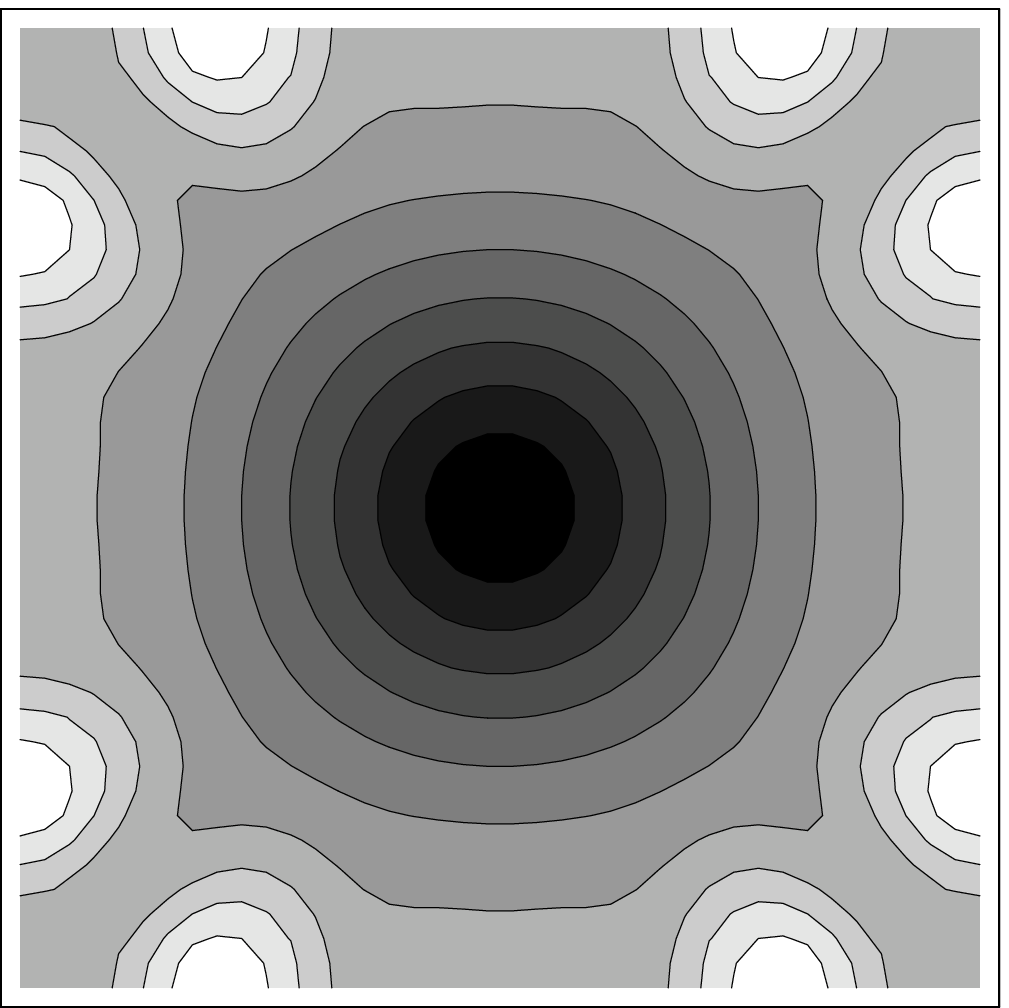}
      }
      \put(0.2,-0.3){(a)}
      \put(2.4,-0.3){(b)}
      \put(4.65,-0.3){(c)}
   \end {picture}
   \end {center}
   \caption
       {%
       A contour plot of $\Omegap$ vs. $p_x$ and $p_y$ for
       $p_z = 0.1$ for eqs.~(a) \protect\ref{eq:spikeless 1},
       (b) \protect\ref{eq:spikeless 2}, and (c) \protect\ref{eq:spikeless 3}.
       \label{fig:nospike}
       }%
   }%
\end {figure}

A simple example of an $\Omegap$ that avoids spikes is
\begin {equation}
   [\Omegap^{(1)}]^2 =
   4 - 2 \prod_i ( 1 + \cos p_i ) \,.
\label{eq:spikeless 1}
\end {equation}
This example is a bit degenerate: the boundary of the Brillouin zone
$|\p_i| \le \pi$ is a constant energy surface, $\Omegap^2 = 4$.
A contour plot of $\Omegap$ for fixed $p_z$ is shown in
fig.~\ref{fig:nospike}a.
There is a maximum along planes of symmetry, but it is a degenerate
maximum.  The nominal conditions for a singularity in the damping
rate are met but only at the boundary, and at the boundary the
group velocity $\v$ corresponding to (\ref{eq:spikeless 1}) vanishes.
The explicit factors of $v^i$ and $v^j$ in the formula
(\ref{eq:gamma}) for $\gamma$ then compensate the divergence of the
$\delta(\v\cdot\k)$, and there are no singularities.

Here's an example that's degenerate in a different way:
\begin {equation}
   [\Omegap^{(2)}]^2 =
   \left\{ (1 - \cos p_x) \, (\cos p_y - \cos p_z)^2 \right\}
   \times
   \left\{ \hbox{cyclic permutations} \right\} \,.
\label{eq:spikeless 2}
\end {equation}
This example has been constructed to give $\Omegap = 0$ along all the
planes of symmetry, and a fixed $p_z$ slice is shown in
fig.~\ref{fig:nospike}b.
This example by itself doesn't have the right $\p \to 0$ behavior for the
continuum theory, but we can now obtain an interesting non-degenerate
example by an appropriate superposition of the two degenerate ones.
For example, a $p_z$ slice of
\begin {equation}
   \Omegap^2 = [\Omegap^{(1)}]^2 + 5\, [\Omegap^{(2)}]^2
\label {eq:spikeless 3}
\end {equation}
is plotted in fig.~\ref{fig:nospike}c.


\section {The plasma frequency and Debye mass}
\label {sec:plasma freq}

Since all the machinery was put in place in section~\ref{sec:review},
it's interesting to look at some other quantities, characteristic of
classical thermal gauge theories, that are not as directly related to
the topological transition rate as the damping coefficient $\gamma$.
In this section, I shall examine the plasma frequency and Debye
screening length at leading order in weak coupling.

The plasma frequency $m_{\rm pl}$
is the frequency of small-amplitude, propagating,
colored waves in the long wavelength limit.\footnote{
   This concept certainly makes sense perturbatively.
   It is not clear (to me at least)
   whether there is a precise, useful, gauge-invariant,
   non-perturbative definition
   of the plasma frequency in non-Abelian gauge theory.
}
(In quantum theories, it is also known as the plasmon mass.)
It can be extracted from the general hard-loop equation
(\ref{eq:blaizot}) by linearizing in $A^\mu$ and taking the $\k \to 0$
limit for fixed $\omega$.  It's convenient to work in a covariant
gauge, where this limit gives
\begin {equation}
   \partial_\mu \partial^\mu A^i =
   - 2 \CA g^2 \int_\p \dn \, v^i v^j A^j \,.
\end {equation}
If the underlying theory has at least cubic symmetry, this can be
rewritten as
\begin {equation}
   \partial_\mu \partial^\mu A^i = m_{\rm pl}^2 A^i \,,
\end {equation}
where
\begin {equation}
   m_{\rm pl}^2 = - {\textstyle{2\over3}} \CA g^2 \int_\p \dn \, |\v|^2 \,.
\end {equation}
For a classical theory, this is
\begin {equation}
   m_{\rm pl}^2 = {\textstyle{2\over3}} \CA g^2 T \int_\p {|\v|^2\over\Omega^2}
   \,.
\end {equation}
For the Kogut-Susskind Hamiltonian on a simple cubic lattice, the
dispersion relationship (\ref{eq:SC omega}) gives
\begin {equation}
   m_{\rm pl}^2 = 0.08606 \; \CA g^2
\label {eq:plasmon}
\end {equation}
in lattice units.

The plasma frequency is in fact related to the physics of topological
transitions.  When the system crosses the potential energy
barrier for such transitions, it oscillates many times back and forth
across the barrier for each net transition \cite{alpha5}.  These oscillations
are small amplitude, and their oscillation frequency is the plasma frequency.
For comparison, a typical SU(2) simulation by Ambj{\o}rn and Krasnitz
\cite{ambjorn} has couplings given by $\beta_{\rm L} \equiv 4/g^2 = 12$, for
which (\ref{eq:plasmon}) predicts a plasma oscillation period of $2\pi/m_{\rm
pl} = 26.2$.  Their simulation has time steps of $0.05$ in lattice units.

Debye screening is the screening of static electric fields in a plasma.
It can be seen from the small-amplitude, zero frequency behavior of
(\ref{eq:blaizot}), which gives
\begin {equation}
   \nabla^2 A^0 = \md^2  A^0 \,,
\end {equation}
where
\begin {equation}
   \md^2 = - 2 \CA g^2 \int_p \dn \,.
\end {equation}
For a classical theory,
\begin {equation}
   \md^2 = 2 \CA g^2 T \int_p {1\over\Omegap^2}  \,.
\label{eq:debye cl}
\end {equation}
For the Kogut-Susskind Hamiltonian on a simple cubic lattice, this gives
\begin {equation}
   \md^2 = 0.50546 \; \CA g^2
\label {eq:debye}
\end {equation}
in lattice units.
In fact, there is an analytic result \cite{Sigma}%
\footnote{
   $\int_p \Omegap^{-2}$ is $\Sigma/4\pi$ where $\Sigma$ is given by
   eq.\ (A.5) of ref.\ \cite{Sigma}, except that equation has a
   typographic error: the elliptic function $K$ should be squared
   \cite{private}.
}
for the integral,
and (\ref{eq:debye}) is more precisely
\begin {equation}
   \md^2 = {4\over\pi^2} \; 
           (18 + 12\sqrt{2} - 10\sqrt{3} - 7\sqrt{6}) \;
           \Bigl[K \Bigl( (2-\sqrt3)^2(\sqrt3-\sqrt2)^2 \Bigr)\Bigr]^2 \;
           \CA g^2 \,,
\end {equation}
where $K(z)$ is the complete elliptic integral of the first kind.

In the weak coupling limit, the inverse screening distance $1/\md$ should
be large compared to the lattice spacing.  For comparison, the weakest
coupling simulations in ref.~\cite{ambjorn} have $\beta_{\rm L}=14$,
for which (\ref{eq:debye}) gives $1/\md = 1.86$ lattice spacings.
So there is one physical quantity, at least, for which these simulations
are only marginally in the weak-coupling limit.
The implications for measuring the weak-coupling behavior of the
topological transition rate are unclear.


\section {Conclusion}

In this paper, I have presented a procedure for extracting the
weak-coupling limit of the real topological transition rate
from rates that could be measured in simulations.
The concerns about rotational invariance originally raised
by Bodeker {\it et al.}\ \cite{bodeker} are crucial,
and a precise measurement of the rate requires a search
for an effectively rotational-invariant lattice Hamiltonian.
Unlike the rotational improvements familiar to Euclidean
studies, simply improving the rotational invariance of higher
and higher derivative terms in the small momentum limit is not
enough: one must also introduce a UV cut-off distance $\Lambda^{-1}$
that is large compared to the lattice spacing, as discussed in
section~\ref{sec:proposal}.

There remain several interesting problems.  One is to explicitly construct
the improved Hamiltonian of section~\ref{sec:proposal}.
Another is to survey other, simpler Hamiltonians (which would not
require numerical extraction of the $a\Lambda \to 0$ limit) to find
one with relatively small anisotropy as measured by
$\sigma_\gamma/\bar\gamma$.  It would also be interesting to find
pure gauge theories that avoid logarithmic spikes, like the
scalar QED theories discussed in sec.~\ref{sec:nospike}.
And, of course, there is the ongoing numerical problem of
actually extracting lattice topological transition rates in the
first place and verifying that they scale as $g_\lat^{10}$ at sufficiently
small coupling.

In this paper, I have focused on extracting the real topological rate
from lattice simulations where the only degrees of freedom are
standard gauge link variables.
An interesting alternative suggested
by Hu and M{\"u}ller \cite{hu&muller} is to instead introduce additional
particle degrees of freedom that are described by continuum positions
$\x(t)$ and classical non-Abelian charges and which interact with
the lattice fields.  The purpose of these particles is to mimic the
hard degrees of freedom of the real quantum theory and induce, in
the lattice theory, the correct long-distance physics of the real
theory.  I would just like to point out that any proposal of this type
will require very careful attention to orders of limits, because there
are in fact {\it two} contributions to the effective long-distance
theory:
that of the new particles, and that of the short-distance modes of
the lattice gauge fields.  The latter gets large as the lattice spacing
is made small and will be anisotropic.
So two non-commuting limits are required of any such scheme:
the limit of small lattice spacing,
and a limit where the coupling to the additional particles is somehow
made large so that their contribution to long-distance physics dominates.

\bigskip

I am particularly indebted to Dam Son for explaining to me Blaizot and Iancu's
formulation of hard thermal loops and for helping me understand how to
generalize it.
I also thank Steve Sharpe, Rajamani Narayan, Greg Moore,
Larry McLerran, Misha Shaposhnikov, and especially Larry Yaffe
for a variety of useful conversations.
This work was supported by the U.S. Department of Energy,
grant DE-FG03-96ER40956.


\appendix

\section {Diagrammatic derivation of \lowercase{$\gamma^{ij}$}}
\label {apndx:diagrammatic}

\begin {figure}
\vbox
   {%
   \begin {center}
      \leavevmode
      
      \epsfbox {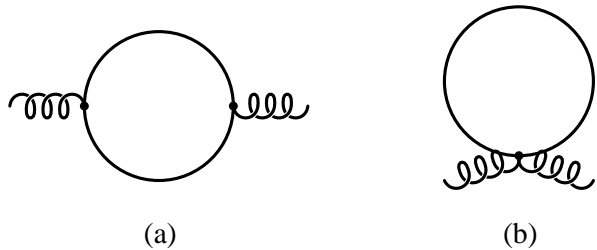}
   \end {center}
   \caption
       {%
         Schematic contributions to the one-loop self-energy of soft
         gauge bosons.  The loop denotes hard excitations of the theory.
       \label{fig:pi}
       }%
   }%
\end {figure}

In this appendix, I will verify the formula (\ref{eq:gamma}) for
$\gamma^{ij}(\k)$---and in particular its interpretation of $\v$ as the group
velocity---by applying diagrammatic methods in a wide class of theories.
Specifically, I will discuss the calculation of the imaginary part $\im\Pi$ of
the gauge boson self-energy starting from one-loop Euclidean diagrams of
the form of fig.~\ref{fig:pi}a, where the line in the loop represents hard
excitations, be they gauge particles, scalars, or whatever.
(Diagrams of the form of fig.~\ref{fig:pi}b will not give a contribution to
$\im\Pi$ because they cannot be cut.)  I will assume that
tree-level propagators of the theory are of the form
$(p_0^2 - \Omegap^2)^{-1}$.
Color indices, which
are just responsible for the overall factor of $\CA$ in
the final result (\ref{eq:gamma}) for $\gamma$, will be ignored.
My first assertion is that the thermal contribution
to $\im\Pi^{ij}$ is
\begin {equation}
   \im\Pi^{ij}(\omega,\k) = - {A_-^{ij}(\omega,\k) \over n^{\rm B}_\omega}
\label{eq:start}
\end {equation}
where $n^{\rm B}$ is the distribution function for bosons;
\begin {equation}
   A_-^{ij}(\omega,\k) = \sum_{ab}
        \int_p {n_\p^{(a)} \over 2\Omegap^{(a)}}
        {(1 \pm n_{\p+\k}^{(b)}) \over 2\Omega_{\p+\k}^{(b)}}
        (\M_{ab}^i)^\star \M_{ab}^j
        \, 2\pi \,
        \delta(\Omega_\p^{(a)} - \Omega_{\p+\k}^{(b)} - \omega) \,,
\end {equation}
is the integrated, squared amplitude for absorption of a gauge boson by
the thermal bath;
and $\M^i = \M^i(K,P)$ is the vertex shown in fig.~\ref{fig:absorb}.
The sum above is over
different species and polarizations $a,b$
of the hard particles, and the capital
letters $K$ and $P$
will henceforth stand for the ``four-vectors'' $(\omega,\k)$
and $(\Omegap,\p)$.
$1 \pm n_\p$ represents the final
state Bose or Fermi enhancement and should be replaced by just $n_\p$ for
a classical theory.

\begin {figure}
\vbox
   {%
   \begin {center}
      \leavevmode
      
      \epsfbox {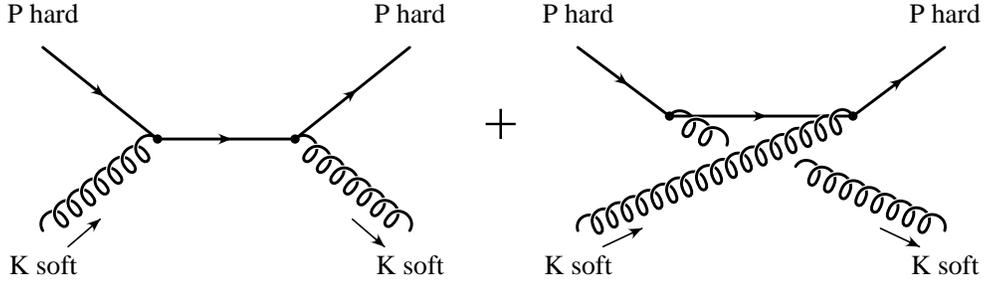}
   \end {center}
   \caption
       {%
         A picture of the thermal part of fig.~\protect\ref{fig:pi}a as forward
         scattering off of particles in the thermal bath.  The
         external hard lines are on shell.
       \label{fig:pi2}
       }%
   }%
\end {figure}

\begin {figure}
\vbox
   {%
   \begin {center}
      \leavevmode
      
      \epsfbox {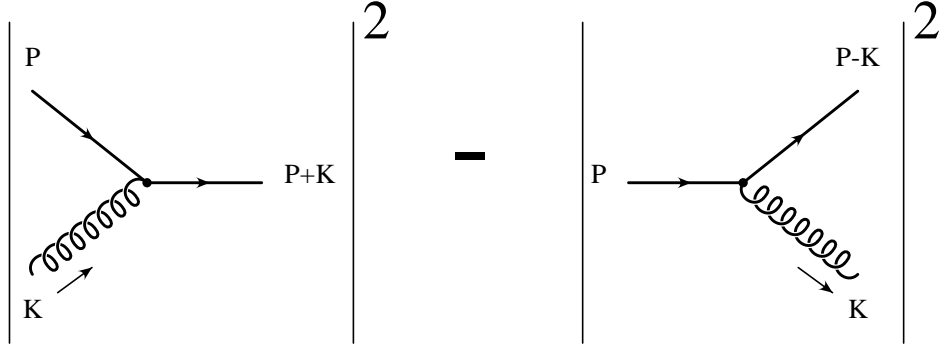}
   \end {center}
   \caption
       {%
         A picture of $-\im\Pi$ as corresponding to the difference
         between absorption and emission processes of soft quanta
         by scattering of hard particles in the plasma.
       \label{fig:impi2}
       }%
   }%
\end {figure}

(\ref{eq:start}) is a simple consequence of the thermal optical theorem
\cite{kobes}.
Rather than plodding through the one-loop case step by step here,
I will give a quick pictorial argument for readers who are unfamiliar
with it.  By the contour
trick for doing Euclidean finite-temperature sums, or by simply starting
directly with a real-time formulation of thermal perturbation theory,
the thermal contribution to the self-energy diagram of fig.~\ref{fig:pi}a
is given by the diagrams of fig.~\ref{fig:pi2},
which represent forward scattering of the
soft excitation off of other excitations in the thermal bath.
The imaginary part corresponds to cutting these diagrams and putting
the cut line on shell.  For the retarded self-energy, where $\omega$
is treated as $\omega + i\epsilon$, this cutting yields the
net decay rate (actually, the rate times $-2\omega$) of excitations:
$-\im\Pi$ is the {\it difference}
$A_- - A_+$ shown in fig.~\ref{fig:impi2} of the square amplitudes $A_-$
for destroying soft quanta and $A_+$ for creating them. 
By the principle of detailed
balance (which is indeed obeyed diagrammatically),
these amplitudes must satisfy
\begin {equation}
   n_\omega A_+ = A_- (1 + n_\omega) \,,
\end {equation}
from which one then has
\begin {equation}
   \im\Pi = -(A_- - A_+) = - {A_- \over n_\omega} \,.
\end {equation}

So return to (\ref{eq:start}).
Taking the limit that both $\omega$ and $\k$ are small and the limit that
$\omega \ll k$ gives
\begin {eqnarray}
   \im\Pi^{ij} &\simeq&
   - 2\pi
   \beta\omega \sum_{\Omega^{(a)} = \Omega^{(b)}} \int_p
      n_\p (1 \pm n_\p) \,
      {(\M_{ab}^i)^\star \M_{ab}^j \over 4 \Omegap^2} \,
      \delta(\v_a\cdot\k)
\nonumber\\
  &=&
   2\pi\omega \sum_{\Omega^{(a)} = \Omega^{(b)}} \int_p
      {dn\over d\Omega} \,
      {(\M_{ab}^i)^\star \M_{ab}^j \over 4 \Omegap^2} \,
      \delta(\v_a\cdot\k)
\label {eq:almost there}
\end {eqnarray}
The sum over flavors is now restricted to pairs of flavors that
have degenerate dispersion relationships.
The result (\ref{eq:gamma}), or its generalization to other theories,
will now follow if $\M_{ab}$ can be
replaced by $g \nabla (\Omegap^2) \delta_{ab} = 2 g \Omegap \v_a \delta_{ab}$:
that is, if
soft magnetic gauge bosons indeed couple to group velocity.
I'll argue this by surveying a few instructive examples.

As the first class of theories to check, start with
a continuum scalar QED theory with a Lagrangian with
scalar interactions of the form
\begin {equation}
   {\cal L}_{\rm scalar}
        = (\partial_t\phi_a)^* (\partial_t\phi_a)
           - \phi_a^* \, f_{ab}(\D) \, \phi_b \,,
\label {eq:scalar toy}
\end {equation}
where $a$ and $b$ are flavor indices.  For the one flavor case,
everything is trivial.  The dispersion relationship is
$\Omegap^2 = f(\p)$, and the trilinear coupling with the photon is
$g \phi^* \phi \A \cdot \nabla f$; so $\M$ has the desired form.

The case of multiple flavors is slightly less trivial and is worth
understanding because it is analogous to the case of multiple branches
in lattice non-Abelian gauge theory.  Let $U(\p)$ be the unitary 
matrix that diagonalizes $f(\p)$ in flavor space, so that
$F = U^\dagger f U$ is diagonal.  It's diagonal elements $F_a$ are just the
$\Omegap^2$ for the different branches.  In this energy eigenbasis,
the trilinear coupling is not in general diagonal:
\begin {equation}
    \M = U^\dagger (\nabla f) U
      = \nabla F + U^\dagger (\nabla U) F + F (\nabla U^\dagger) U
      = \nabla F + [ U^\dagger \nabla U, F ] \,,
\end {equation}
so that
\begin {equation}
   \M_{ab} = \nabla F_a \delta_{ab} + (U^\dagger \nabla U)_{ab} (F_b - F_a)
   \,.
\end {equation}
But the degeneracy condition on the flavor sum in (\ref{eq:almost there})
eliminates the second term above, and so one obtains
\begin {equation}
   \gamma^{ij}(\k) = - 2 \pi \sum_{a} g^2
       \int_\p \dn \, v_a^i v_a^j \, \delta(\v_a \cdot \k) \,.
\end {equation}
This is indeed the trivial generalization of (\ref{eq:gamma}) to
include a sum over branches.

Now consider scalar QED on a spatial lattice, with the gradient energy
given by some sum of local terms of the form
\begin {equation}
   \left|\phi(\x) - U_{\x\y} \phi(\y)\right|^2 \,,
\end {equation}
where $U_{\x\y}$ is the product of U(1) link matrices along some path
from $\x$ to $\y$.  (The canonical case would be that $\x$ and $\y$ are
nearest neighbors and the path is the single link between them.)
I am only interested in the $\phi^* A \phi$ coupling
in perturbation theory and only in the limit of soft momentum
for $A$.  In this limit, $U_{\x\y} \approx 1 + g (\x-\y)\cdot\A$ and
the coupling is minimal.  That is, the coupling has the form
(\ref{eq:scalar toy}) in momentum space in the soft photon limit, and the
previous discussion applies.

Similar considerations hold for non-Abelian lattice gauge theories.
Schematically, write $A$ = $A_\soft + A_\hard$ where $A_\soft$ are the
soft-momentum modes and $A_\hard$ the hard momentum modes.
Under {\it soft} gauge transformations $G$, we have
$A_\soft \to G A_\soft G^{-1} - g^{-1} G \nabla G^{-1}$ and
$A_\hard \to G A_\hard G^{-1}$.  Gauge invariance under soft gauge
transformations then implies that the $A_\hard A_\soft A_\hard$
coupling must be minimal in the soft limit.  That is, it again has the
form of (\ref{eq:scalar toy}) in the soft limit with $\phi$ representing
$A_\hard$ and $\D = \nabla - g \A_\soft^a {\cal T}^a$ where ${\cal T}^a$
are the (anti-Hermitian) adjoint-representation generators.

As a check, it is easy to explicitly verify that the
soft-hard-hard coupling of the Kogut-Susskind Hamiltonian on
a simple cubic lattice indeed has $\M_{ab} = g \nabla (\Omegap^2) \delta_{ab}$,
where $a$ and $b$ are the polarizations of the hard gluons.


\section {Numerical integrals for $\tr\,\gamma$}
\label {apndx:numerical}

For the sake of completeness, I will explicitly write down the
two-dimensional integrals corresponding to (\ref{eq:gamma}) that
I did numerically to produce figs.~\ref{fig:impi}b and c.

\subsection {Simple cubic case}

\begin {equation}
   \tr\,\gamma(\hat\k) = {1\over4\pi^2|\k|} \CA g^2
     \int_{-\pi}^{+\pi} {dp_x dp_y} \> \sum_{p_z}
     {|\v|^2 \over \Omegap} \>
     {\theta(1-|\sin p_z|)
           \over \hat k_z |\cos p_z|} \,
     \,,
\end {equation}
where the $p_z$ sum is over two values $p_z^\pm$ with
\begin{mathletters}%
\begin {eqnarray}
   \sin p_z &=&
         - {\hat k_x\over\hat k_z} \sin p_x
         - {\hat k_y\over\hat k_z} \sin p_y \,,
\\
   \cos p_z &=&
         \pm \sqrt{1 - \sin^2 p_z} \,,
\\
   \sin^2\left(p_z\over2\right) &=&
         {\textstyle{1\over2}} (1 - \cos p_z) \,.
\end {eqnarray}%
\end{mathletters}%
$\Omegap$ and $\v$ are given by (\ref{eq:SC omega}) and
(\ref{eq:SC v}).


\subsection {FCC case}

\begin {eqnarray}
   \tr\,\gamma(\hat\k) = {1\over8\pi^2|\k|} \CA g^2
     \int_{-2\pi}^{+2\pi} dp_x dp_y && \sum_{\Omega_\pm} \sum_{p_z}
     {|\v_\pm|^2 \over \Omega_\pm} \>
     {2\sqrt{1-{1\over16}\epsilon_\p} \over
       \left|B \cos\left(p_z\over2\right)
          - A\sin\left(p_z\over2\right)\right|} \>
\nonumber\\ && \qquad
     \times
     \theta(A^2 + B^2 - C^2) \,
     \theta\left(1-\left|\sin\left(p_z\over2\right)\right|\right)
     ,
\end {eqnarray}
where the $\Omega$ sum is over the two dispersion relations
(\ref{eq:FCC omega});
$\epsilon_\p$ and $\v_\pm$ are given by (\ref{eq:FCC epsilon})
and (\ref{eq:FCC v}) with $\Omegap = \Omega_{\pm}$; the
$p_z$ sum is over two values $p_z^\pm$ with
\begin {mathletters}%
\label {eq:FCC pz}%
\begin {eqnarray}
   \sin\left(p_z\over2\right) &=&
        { B C \pm A \sqrt{A^2 + B^2 - C^2}
          \over A^2 + B^2 } \,,
\\
   \cos\left(p_z\over2\right) &=&
        { A C \mp B \sqrt{A^2 + B^2 - C^2}
          \over A^2 + B^2 } \,;
\end {eqnarray}%
\end{mathletters}%
and
\begin {eqnarray}
   A &\equiv& \hat k_x \left[
       \sin\left(p_y\over2\right) + \sin\left(p_z\over2\right) \right] \,,
\\
   B &\equiv& \hat k_z \left[
       \cos\left(p_x\over2\right) + \cos\left(p_y\over2\right) \right] \,,
\\
   C &\equiv&
     - \hat k_x \sin\left(p_x\over2\right) \cos\left(p_y\over2\right)
     - \hat k_y \sin\left(p_y\over2\right) \cos\left(p_x\over2\right) \,.
\end {eqnarray}
The $\pm (\mp)$ sign in (\ref{eq:FCC pz}) is unrelated to that in
(\ref{eq:FCC omega}).


\end {document}